\documentclass[%
 reprint, %linenumbers,
nofootinbib,
pra,
]{revtex4-2}

\usepackage{graphicx}% Include figure files
\usepackage{dcolumn}% Align table columns on decimal point
\usepackage{bm}% bold math
\usepackage{amsfonts}

\usepackage{color}
\usepackage{xcolor}

\usepackage[english%,russian
]{babel}
\begin{document}

\title{\textbf{Nonperturbative regime of low-order harmonic generation \\ in intense low-frequency laser field}}% 
\author{
S.~A.~Bondarenko$^{1,2}$
V.~V.~Strelkov$^{1\ast}$
}
\affiliation{
\mbox{$^{1}$P.N. Lebedev Physical Institute of the Russian Academy of Sciences, 53 Leninskiy Prospekt,  Moscow 119991, Russia} \\
\mbox{$^{2}$National Research Nuclear University MEPhI, 31 Kashirskoe Highway, Moscow 115409, Russia} \\
%\mbox{$^{3}$A. V. Gaponov-Grekhov Institute of Applied Physics of the Russian Academy of Sciences,} \\ {46 Ulyanov street, Nizhny Novgorod 603950,	Russia} \\
$^{\ast}$strelkov.v@gmail.com
}

\date{\today}% It is always \today, today,
             %  but any date may be explicitly specified

\begin{abstract}
%We find the atomic response to an intense femtosecond laser pulse via solving numerically the three-dimensional non-stationary Schr\"odinger equation (TDSE) for a model atom and calculating its dipole moment. For weak quasi-static fields, the response is well described by a perturbation approach, but for intensities higher than a certain value (about $0.6 \,  \,  10^{14}$~W/cm$^2$ for argon) the accuracy of this description is unsatisfactory, regardless of the order of non-linearity taken into account. We suggest fitting the numerical TDSE solution results with a Pad\'e expansion, and show that this approximation describes the response well both in the perturbative regime and beyond it for intensities up to approximately half of that marking the onset of barrier-suppression ionization (in particular, for argon the approximation is applicable up to $1.4 \,  \,   10^{14}$ W/cm$^2$). To consider the non-perturbative nonlinearity beyond the quasi-static limit we use the model of nonlinear oscillator with the restoring force defined by the found Pad\'e expression. Our model fails to predict the behaviour of the nonlinear refractive index in the nonperturbative domain, but it describes well the nonperturbative growth of the efficiency with the laser intensity for other nonlinear optical processes, namely, the third and fifth harmonic generation in the IR field and the optical rectification in a two-color field.

We find the atomic response to an intense femtosecond laser pulse via solving numerically the three-dimensional non-stationary Schr\"odinger equation (TDSE) for a model atom and calculating its dipole moment. For weak quasi-static fields, the response is well described by a perturbation approach, but for intensities higher than  a certain value (about $0.6 \,  \,  10^{14}$~W/cm$^2$ for argon) the accuracy of this description is unsatisfactory, regardless of the order of non-linearity taken into account. We suggest fitting the numerical TDSE solution results with a Pad\'e expansion, and show that this approximation describes the response well both in the perturbative regime and beyond it  for intensities up to approximately half of that marking the onset of barrier-suppression ionization (in particular, for argon the approximation is applicable up to $1.4 \,  \,   10^{14}$ W/cm$^2$). To consider the non-perturbative nonlinearity beyond the quasi-static limit we use the model of nonlinear oscillator with the restoring force defined by the found Pad\'e expression. Our model fails to predict the behaviour of the nonlinear refractive index in the nonperturbative domain, but it describes well the nonperturbative growth of the efficiency with the laser intensity for other nonlinear optical processes, namely, the third and fifth harmonic generation in the IR field and the optical rectification in a two-color field.
\end{abstract}
                             
\maketitle

\section*{Introduction}
Harmonic generation in the laser fields is one of the fundamental phenomena in non-linear optics. For moderate laser intensities and low harmonic orders, this process can be successfully described with a perturbation theory~\cite{Shen}. For high intensities (higher than $10^{13} - 10^{14}$~W/cm$^2$) which can be used for harmonic generation in gases, photoionization of atoms leads to non-perturbative  mechanisms of low-~\cite{Brunel} and high-~\cite{3-step_C} order harmonic generation (for reviews see~\cite{Xiong_2017, Ryabikin2023}).  Such processes have been successfully studied analytically, first of all, using the strong-field approximation~\cite{Becker_1994, Lewenstein, EVA}. However, the non-perturbative properties of low- and moderate-harmonic generation appear even for lower fields, when these properties are not directly linked to the photoionization.  This behaviour is actively investigated using Floquet approach~\cite{Frolov_2007, Cheng_2018, Antonov2025}, as well as via analysing the numerical results, see Refs.~\cite{ABecker2014, ABecker2015, ABecker2017,  Emelin2025}.

  Low- and moderate-order harmonic generation has recently found significant applications in precision spectroscopy and ultrafast metrology. In particular, in~\cite{Zhang_2024} they used low-order harmonics to measure the nuclear transition frequency of Thorium-229, with important implications for future nuclear clocks. In~\cite{Han_2025} chiral asymmetries in the forward-backward and angle-resolved photoionization delays of chiral molecules were studied using elliptically polarized low-order harmonics. Moreover, recent works~\cite{Wang_2026, Han_2026} have explored low-order harmonic generation for ultrafast chiral sensing and time-resolved spectroscopy.

In a periodic field $E(t)=E_0 \exp(-i \omega_0 t)+c.c.$ the microscopic response spectrum $d(\omega)$ consists of harmonics. For the moderate intensity of the field for the non-resonant case the response can be described within the perturbation theory~\cite{Shen} and the $q$-harmonic spectral amplitude is related to the field amplitude as 
\begin{equation}
d(q \omega_0)=\tilde \alpha(q \omega_0)E_0^q,
\label{alpha}
\end{equation}
here the effective (field's amplitude dependent) non-linear polarizability $\tilde \alpha(q \omega_0)$ is
\begin{equation}
\begin{array}{cc}
 
    & \tilde \alpha(q \omega_0)= \\
     & \sum_{j=0,2,...} \alpha^{(q+j)} (q \omega_0=\underbrace{\omega_0+...+\omega_0}_{q+j/2}\underbrace{-\omega_0...-\omega_0}_{j/2}) E_0^{j}
    \end{array} 
    \label{alpha_exp} 
\end{equation}
where $\alpha^{(q+j)} (q \omega_0)$ is the nonlinear polarizability characterizing the $(q+j)$-th order contribution to the $q$-th harmonic generation. This polarizability does not depend on the field amplitude. Note that only odd $q$ and $q+j$ terms are non-zero in an inverse-symmetric system. Certainly, for relatively weak fields the first term in this series dominates, so $\tilde \alpha(q \omega_0) \approx \alpha^{(q)} (q \omega_0)$. Note also that in this paper we deal with the microscopic response, so we characterize it with the polarizability; commonly used in nonlinear optics susceptibility $\chi^{(q)}$ characterizes strictly speaking the {\it macroscopic} response, and that is why susceptibility (though clearly related to polarizability) is not discussed in this paper.

The expansion~(\ref{alpha_exp}) is applicable if this series converges. This is the case when 
\begin{equation}
 E_0^2 \,  \alpha^{(q+2)} \ll \alpha^{(q)}.
 \label{conv}
\end{equation}
Thus, within perturbation theory $\tilde \alpha(q \omega_0)$ in Eq.~(\ref{alpha}) depends (i) weakly and (ii) almost linearly on the laser intensity. For $q=1$ this dependence is well-known as the Kerr-effect. Note that as the intensity increases, pronounced deviation from the linear dependence can take place. This means that requirement~(\ref{conv}) is not fulfilled, so the perturbation series does not converge. Under such conditions, the perturbation approach can not be applied. So, the deviation from this linear dependence on the intensity~\cite{Mechain_2004, Loriot_2009, Bejot_2010, Ettoumi_2010, Kosareva_2011, Popov_2013, volkova_nonlinear_2013, Shipilo_2017} should be attributed rather to the non-perturbative effects~\cite{volkova_nonlinear_2013, ABecker2015} than to the higher-order non-linearities.   

When the field's frequency $\omega_0$ is low enough so that the microscopic response $d$ depends only on the instantaneous strength of the field $E$, the response as a function of the field can be expanded in the Taylor series:

\begin{equation}
    d(E) = \sum_{m=1, 3 \dots} \kappa^{(m)} E^{m}
    \label{eq:series}
\end{equation}

In this case using the binomial expansion $\left[ E(t)\right]^m= E_0^m\left[\left(\exp(-i \omega_0 t)+ \exp(i \omega_0 t)\right)/2\right]^m=E_0^m 2^{-m} \sum_{k=0}^m C^k_m \exp\left(-i \omega_0 t [m-2k]\right)+c.c.$ (where $C^k_m=m!/(k!(m-k)!)$ is a binomial coefficient), one can connect nonlinear polarizabilities  $\alpha^{(q+j)} (q \omega_0)$ and $\kappa^{(q+j)}$ coefficients: 
\begin{equation}
 \alpha^{(q+j)} (q \omega_0)=\kappa^{(q+j)}\frac{(q+j)!}{2^{q+j-1}(j/2)!(q+j/2)!}, \, \, \, j=0,2,...
\end{equation}

Several lowest-order coefficients are:
\begin{equation}
   \begin{array}{cc}
     &  \alpha^{(1)} (\omega_0) = \kappa^{(1)}\\
     &  \alpha^{(3)} (\omega_0=\omega_0+\omega_0-\omega_0) = (3/4) \kappa^{(3)}\\
     &  \alpha^{(5)} (\omega_0=\omega_0+\omega_0+\omega_0-\omega_0-\omega_0) = (10/16) \kappa^{(5)}\\
     &  \alpha^{(3)} (3 \omega_0=\omega_0+\omega_0+\omega_0) = (1/4)\kappa^{(3)}\\
     &  \alpha^{(5)} (3\omega_0=\omega_0+\omega_0+\omega_0+\omega_0-\omega_0) = (5/16) \kappa^{(5)}\\
     &  \alpha^{(5)} (5\omega_0=\omega_0+...+\omega_0) = (1/16) \kappa^{(5)}\\
\end{array} 
\end{equation}

Note that coefficients $\kappa^{(m)}$ in Eq.~(\ref{eq:series}) are real, so the non-linear polarizabilities are also real.  

A necessary (but not sufficient, see below) condition of the applicability of the perturbation expansion~(\ref{alpha})~-~(\ref{alpha_exp}) is~\cite{Elutin}:
\begin{equation}
 \Omega_{R} \ll \Delta    
 \label{perturb}
\end{equation}
where $\Omega_R= d_{i,f} E_0 $ is the Rabi frequency, $\Delta = |\omega_0-\omega_{i,f}|$ is the detuning from the resonance, $d_{i,f}$ is the dipole matrix element of the transition from the initial state $i$ to the final state $f$, $\omega_{i,f}$ is the transition frequency.The initial state is usually the ground state. Condition~(\ref{perturb}) should be satisfied for all final states. Moreover, it should be satisfied taking into account the uncertainty of $\Delta$ due to the limited duration of the laser pulse. One can see that this condition is satisfied for weak (low $E_0$) and non-resonant (high $\Delta$) fields. 

The {\it ab initio} numerical studies allow defining the intensities starting from which the nonlinear polarizabilities behave non-perturbatively, demonstrating saturation of the Kerr-effect~\cite{Ettoumi_2010, Kosareva_2011, volkova_nonlinear_2013, Shipilo_2017}, nonlinear dependence of the third-order polarizability on the laser intensity~\cite{volkova_nonlinear_2013} and the dependence of the polarizabilities on the pulse duration~\cite{ABecker2017}. These studies show that under certain conditions non-perturbative properties are observed under the intensities for which (\ref{perturb}) is still satisfied. This could be partly attributed to multiphoton resonances, which naturally have lower detunings and can contribute sufficiently to the polarization of the system already at moderate intensities~\cite{Smetanin_2016}.   However, the response to the quasi-static field can not be attributed to (one- or multi-photon) resonances. So, the nonperturbative behaviour in the quasi-static field before the onset of the ionization requires studying in more detail. 

In this paper, we study numerically the atomic response to the quasi-static electric field. Solving TDSE for a model atom in an external field, we calculate the dipole moment as a function of the field strength. We show that the polynomial series~(\ref{eq:series}) derived for weak intensities is not applicable for the higher ones, and its applicability can not be improved by including higher terms.  
 
We find an alternative approximation for the dipole moment, which describes well the numerical results both for low and high intensities including those for which the series~(\ref{eq:series}) is inapplicable. The applicability range of our approximation is limited by the intensity at which photoionization starts to essentially contribute to the dipole moment of the system.  We apply this approximation to describe the third and fifth harmonic generation, as well as quasi-static field production via two-colour field. 

Moreover, we find up to which laser frequency our quasi-static field approach is applicable. First, we compare the quasi-static results with the numerical one and find the frequencies starting from which the difference is non-negligible. Second, taking into account that the classical linear oscillator model is known to provide reasonable description of the dispersion of the linear polarization, we develop the nonlinear oscillator (known as Duffing oscillator for a cubic nonlinearity) model to describe the dispersion of the nonlinear polarization including the region of intensities where it behaves non-perturbatively. 

\section{Time-dependent Schr\"odinger equation}
To find the atomic response we integrate  numerically 3D time-dependent Schr\"odinger equation (TDSE) for an atom in linearly-polarized external field. TDSE in the single-active electron approximation is written in cylindrical coordinates using atomic units as: 

\begin{equation}
    i\frac{\partial}{\partial t} \Psi(t, \rho, z) = \hat{H}(t) \Psi(t, \rho, z),
\label{eq: tdse}
\end{equation}
where $\Psi(t, \rho, z)$ is the time-dependent wave function of the electron and $\hat{H}(t)$ is the Hamiltonian of the system.

The Hamiltonian is given by:
\begin{equation}
    \hat{H}(t) = -\frac{1}{2} \Delta + U(r) + zE(t) - iW(\rho, z),
    \label{eq: hamiltonian}
\end{equation}
where $r^2=\rho^2+z^2$, $U(r)$ is the effective one-electron potential which models the argon atom and
 $-iW(\rho, z)$ is an imaginary (absorbing) potential (see Appendix A for more details). 

The time-dependent electric field $E(t)$ corresponds to a laser pulse with frequency $\omega_0$ and peak amplitude $E_0$, modulated by an envelope function $g(t)$:
\begin{equation}
    E(t) = E_0 g(t) \cos\left[\omega_0 \left(t - \frac{t_{\text{full}}}{2}\right)\right],
    \label{eq: electric_field}
\end{equation}
where $t_{\text{full}} = N \frac{2\pi}{\omega_0}$ is the full duration of the pulse, $N$ is the number of optical cycles in the pulse, $g(t)=\sin^2 \left[ \frac{\pi t}{t_{full}} \right]$ is a smooth temporal envelope which provides adiabatic turn-on and turn-off of the pulse.

To find the atomic response we calculate the expectation value of the time-dependent dipole moment
\begin{equation}
d(t)=
\int z |\Psi(t,\rho,z)|^2 2\pi \rho d \rho \, dz 
\label{eq:d(t)}
\end{equation}
then we calculate its spectrum $d(\omega)$. 

The effective $q$-th order nonlinear polarizability $\tilde\alpha^{(q)}(\omega)$ is calculated as:
\begin{equation}
     \tilde\alpha^{(q)}(q\omega_0) = \frac{\bar d(q\omega_0)}{\bar E(\omega_0)^q}, 
\label{chi}     
\end{equation}
where 
\begin{equation}
 \bar E(\omega_0)=\left[\frac{\int _{\omega_0-\Delta} ^{\omega_0+\Delta} |E(\omega)|^2 d \omega}{\tau}\right]^{1/2},
\end{equation}
$\tau$ is the FWHM duration of the laser pulse, $\Delta=\omega_0/2$
\begin{equation}
\bar d(q\omega_0) = \left[ \frac{\int _{q\omega_0-\Delta} ^{q\omega_0+\Delta} |d(\omega)|^2 d \omega}{\tau/\sqrt{q}} \right]^{1/2},
\end{equation}
here we take into account that the FWHM duration of the $q$-th harmonic pulse is $\tau/\sqrt{q}$. Note that $\bar E(\omega_0)$ is close to the maximal amplitude of the field $ E_0$, and $\bar d(q\omega_0)$ is close to the maximum amplitude of the harmonic response, that is why these values are used in Eq.~(\ref{chi}) to reconstruct $\tilde\alpha^{(q)}(q\omega_0)$. A simpler way to calculate $\tilde\alpha^{(q)}(q\omega_0)$ would be using a flat-top laser pulse; in this case $\tilde\alpha^{(q)}(q\omega_0) = \frac{ d(q\omega_0)}{E(\omega_0)^q}$. However, our laser pulse given by Eq.~(\ref{eq: electric_field}) is more realistic.  

\section{Response to the quasi-static field.}

For weak quasi-static fields the response as a function of the field can be described by the Taylor expansion~(\ref{eq:series}). In this section, we shall study the region of applicability of the Taylor series as an approximation of the numerical results. Moreover, we shall propose another approximation for describing the response outside of this region, i.e. in the non-perturbative domain.

\subsection{Numerical response}
\label{num_q-s}
We solve TDSE numerically for a few-cycle pulse having very low-frequency (namely, 1.7~10$^{-3}$ a.u.). Since we are focusing at the atomic response, we suppress the contribution of free electrons to the system's polarizability. To achieve this we, first, make the simulation using relatively small numerical box (its boundaries in the field's polarization direction are at $\pm 20$ a.u.), so that the detached part of the electronic wave-packet is absorbed by the box's boundaries and does not contribute to the dipole moment. Second, we correct the response taking into account the depletion of the wave-function due to the photoionization. Namely, instead of Eq.~(\ref{eq:d(t)}) we use:
\begin{equation}
d(t)=
\frac{\int  z |\Psi(t,\rho,z)|^2 2\pi \rho d \rho \, dz}{\int|\Psi(t,\rho,z)|^2 2\pi \rho d \rho \, dz}.
\label{eq:d_corrected(t)}
\end{equation}

For moderate field intensities this response adiabatically traces the laser field, so using it one can directly reconstruct the response to the quasi-static field $d_{qs}(E)$. For higher intensities the photoionization affects the results, so below we present $d_{qs}(E)$ for field amplitudes up to $E=0.063$ a.u. (which corresponds to the intensity 1.4~10$^{14}$~W/cm$^2$). Fig.~\ref{fig:num_app_response} shows the nonlinear part of this response.

\subsection{Polynomial series approximation of the numerical response}
\begin{figure}
    \centering
    \includegraphics[width=0.9\linewidth]{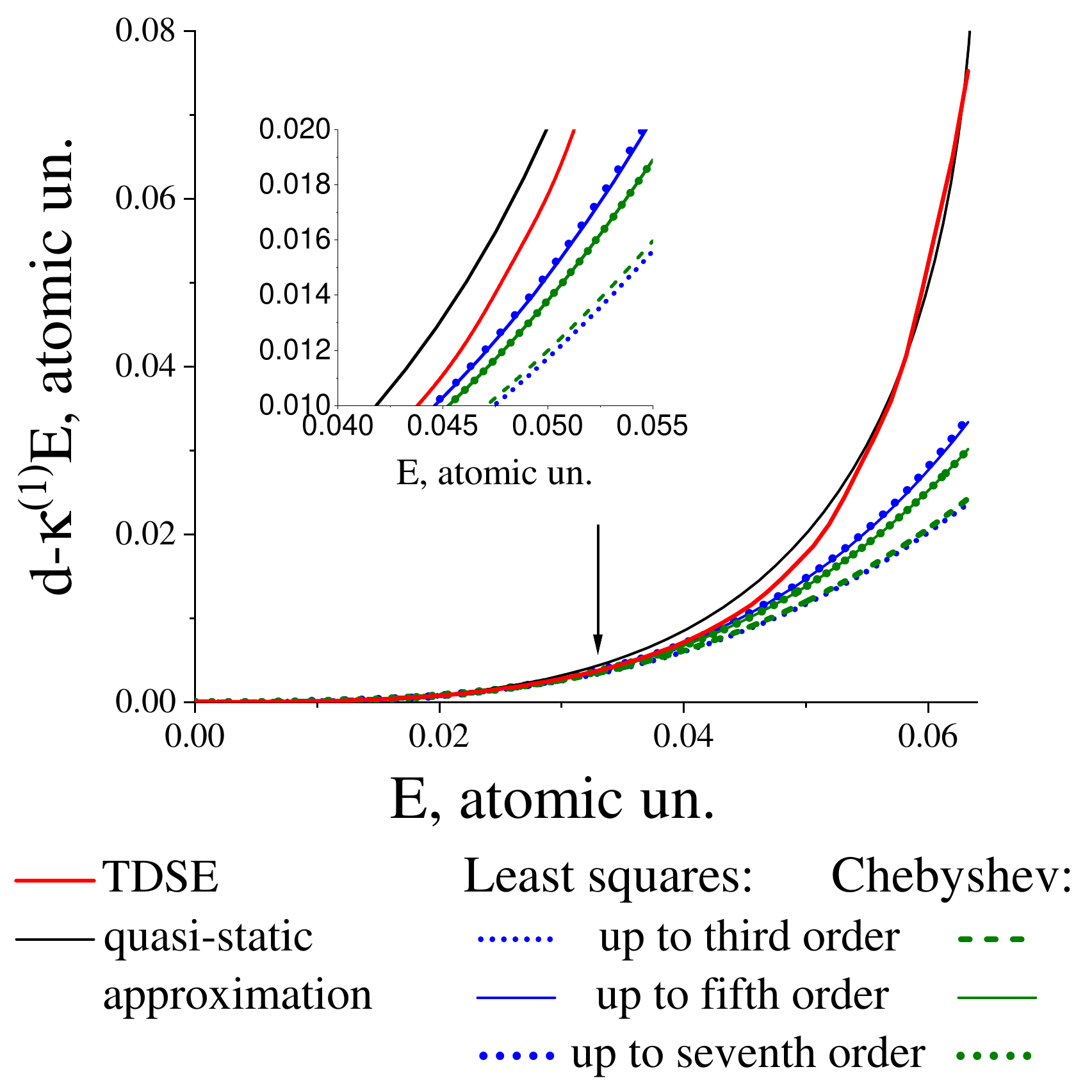}
    \caption{The atomic dipole moment  of the argon atom in the quasi-static field calculated via numerical TDSE solution (solid red curve), its approximation with Pad\'e expansion (solid black curve), and with the polynomial series (see the legend). The black arrow shows the boundary of the approximation interval $E_{max}$. The inset shows the region where the polynomial approximation starts to deviate from the TDSE result, and this discrepancy can not be removed via including higher-order terms in the series.}
    \label{fig:num_app_response}
\end{figure}

One can approximate the response of an atom to the quasi-static field near $E=0$ by a polynomial~(\ref{eq:series}). The well-known form of this approximation is the Taylor series, where the coefficient $\kappa^{(q)}$ is defined by the $q$-th derivative at $E=0$. However, the result of the numerical calculation of high order derivatives is very unstable. Here we are using two alternative methods to construct the polynomial approximation: the least squares method and a Chebyshev polynomial approximation. In the least squares method we approximate $d(E)$ with the series~(\ref{eq:series}) and find coefficients which minimize the deviation from the numerical result within some interval [-$E_{\text{max}}$, $E_{\text{max}}$]. The  Chebyshev approach (also approximating in the interval [-$E_{\text{max}}$, $E_{\text{max}}$]) is described in Appendix B.  We chose the least $E_{\text{max}}$ value that provides converging series up to $q=7$ for the field strengths considered in Fig~\ref{fig:num_app_response}; this value is  $E_{\text{max}}=0.033$.

In Fig.~\ref{fig:num_app_response} we see that both series provide similar results. Moreover, when the field strength exceeds approximately $0.04$ (which corresponds to the intensity 0.6~10$^{14}$~W/cm$^2$) both series (though still converging) have low precision, and adding a next order term does not improve it. Taking into account that the perturbation approach results in a polynomial series~(\ref{eq:series}), we conclude that the low precision of the series approximation means inapplicability of the perturbation theory. 

\begin{figure}
    \centering
    \includegraphics[width=0.9\linewidth]{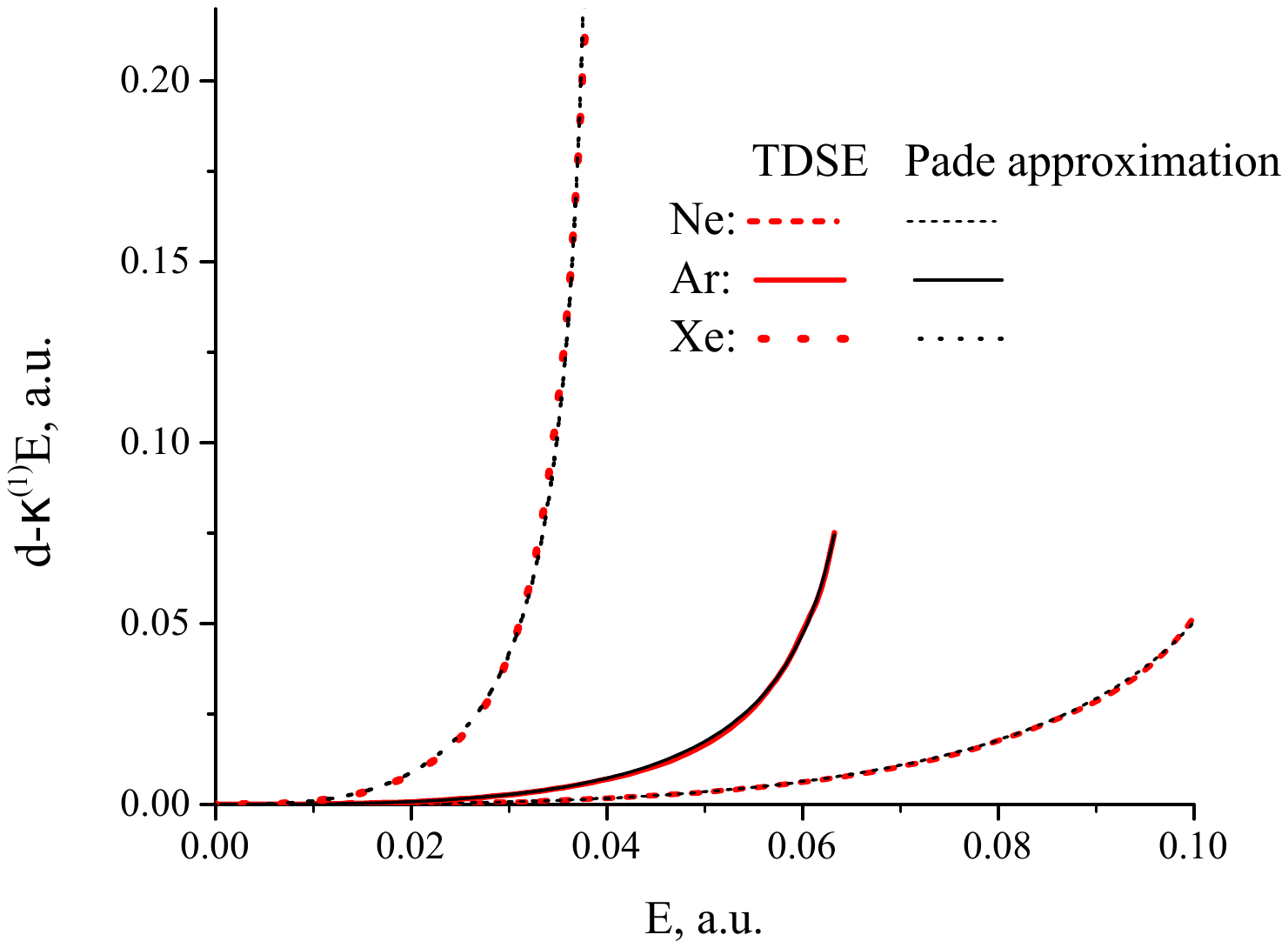}
        \caption{ The atomic dipole moments of neon, argon, and xenon  atoms in the quasi-static field calculated via numerical TDSE solution (red curves) and their approximations with Pad\'e expansion (black curves).}
    \label{fig:num_app_Ne_Ar_Xe}
\end{figure}

\subsection{Pad\'e approximation of the numerical response}

In Fig.~\ref{fig:num_app_response} we see that the numerical response behaves as having singularity at some field strength $E_s$ slightly exceeding the maximal field used in the numerical calculation. So we fit the numerical result with the following Pad\'e approximation:
\begin{equation}
 d_{qs}(E)=\kappa^{(1)} E+\kappa^{(3)}\frac{E}{2}\left( E^2 -E_s^2+\frac{E_s^4}{E_s^2-E^2}\right),
   % E =d_{qs} \frac{b+c\cdot  d_{qs}^4 }{(a^2 + d_{qs}^2)^{3/2}},
\label{eq:Pade}
\end{equation}
where the field $E_s$ is chosen to reproduce the numerical response. The found value of $E_s=0.07$ is close to the field strength for which the Coulomb barrier is suppressed~\cite{Delone_1998}: $E_{BSI}=I_p^2/4$, where $I_p$ is the ionization potential, for argon $E_{BSI}=0.084$.  

 We have chosen the Pad\'e approximation of the response in the form~(\ref{eq:Pade}) because it has the following features:

  - for $E \ll E_s$ the linear and cubic nonlinearities are the same as in the perturbation approach: $d_{qs}(E)=\kappa^{(1)} E+\kappa^{(3)}E^3/2$

 - the singularity for $E \sim   E_s$ contains  the lowest order of $E$ which does not cause (unphysical) even harmonics.

A better approximation for $d_{qs}$ can possibly be found using Siegert Pseudo-States of an atom in a static field~\cite{Tolstikhin_2013}.  Moreover, multielectron effects can be taken into account calculating the atomic response with density functional theory~\cite{DFT, Telnov_2013, Romanov_2024}, in particular, using optimized effective potentials~\cite{Hirata_2002, Yang_2002}. This can be a natural perspective for the future studies. 

 Using model potentials for neon and xenon (see Appendix A for details) we solve TDSE for these atoms in a quasi-static field and calculate the responses. We choose these atoms because they are widely used in strong-field experiments and because they have very different ionization potentials. The comparison of the numerical responses with the corresponding Pad\'e approximations is presented in Fig.~\ref{fig:num_app_Ne_Ar_Xe}.  In all the cases the calculated $F_s$ is close to $E_{BSI}$ for the corresponding atom, namely, we use $E_s=0.78 E_{BSI}$ for neon, $E_s=0.83 E_{BSI}$ for argon and $E_s=0.81 E_{BSI}$ for xenon. Figs.~\ref{fig:num_app_response} and~\ref{fig:num_app_Ne_Ar_Xe} show that the approximation~(\ref{eq:Pade}) agrees well with the TDSE results for different atoms both in the perturbative and the non-perturbative regions. The agreement is found for neon up to $E\approx 0.64 E_{BSI}$ (thus the field intensity is 2.5 times lower than the barrier-suppression intensity), for argon and xenon up to $E\approx 0.75 E_{BSI}$ (the field intensity is 1.8 times lower than the barrier-suppression intensity). 

\section{Nonlinear oscillator model}

In the next sections, we switch from the quasi-static fields to those oscillating at non-zero (though low) frequencies. The simplest model which describes the dispersion of the linear polarizability considers an atom as a linear oscillator (the Lorentz model). In this section we suggest the nonlinear oscillator model to describe the dispersion of the nonlinear polarizability. 

An oscillator driven by an external field $E(t)$ is described as:
\begin{equation}
    \ddot{d}+\Gamma \dot{d}-F(d) = E(t), 
    \label{eq:oscillator}
 \end{equation}
where $d(t)$ is the displacement from the equilibrium, $\Gamma$ is the damping constant, $F(d)$ is the restoring force. The damping constant defines the natural bandwidth of the oscillator emission. Under the conditions employed in our calculations, the actual atomic emission time $1/\Gamma$ is much longer than the laser pulse duration, so we neglect damping in further calculations. 

Our model is defined by the assumption that in the quasi-static limit the displacement and the external force are linked by Eq.~(\ref{eq:Pade}). Having in mind that for the quasi-static case $F=-E$ we have the following implicit equation for $F(d)$:
\begin{equation}
d=-\kappa^{(1)} F(d)-\kappa^{(3)}\frac{F(d)}{2}\left(F^2(d) -E_s^2+\frac{E_s^4}{E_s^2-F^2(d)}\right),
   % E =d_{qs} \frac{b+c\cdot  d_{qs}^4 }{(a^2 + d_{qs}^2)^{3/2}},
\label{eq:forse_osc}
\end{equation}
Tabulating the latter equation, we find the force numerically as a function of displacement $F(d)$. For weak fields one can keep only the linear term in this dependence leading to the linear oscillator model. Keeping the linear and cubic terms leads to the Duffing oscillator model. 

Below we numerically solve equation~(\ref{eq:oscillator}) with the complete non-perturbative dependence of force on the displacement, defined by Eq.~(\ref{eq:forse_osc}). The calculation is done for zero initial conditions and the driving field~(\ref{eq: electric_field}). 

\section{Results}
In this section, we compare the nonlinear polarizabilities obtained by different methods explained above. Namely, we find the dipole moment within the numerical TDSE solution (Eq.~(\ref{eq:d(t)})), within the quasi-static approach using the Pad\'e approximation (calculating $d_{qs}(E(t))$ via Eq.~(\ref{eq:Pade}) for the field~(\ref{eq: electric_field})) and within the nonlinear oscillator model (solving numerically Eq.~(\ref{eq:oscillator})). Then we calculate the corresponding $q$-th order nonlinear polarizability via Eq.~(\ref{chi}).

\begin{figure}
  %  \centering
    a) 
    
    \includegraphics[width=0.8 \linewidth]{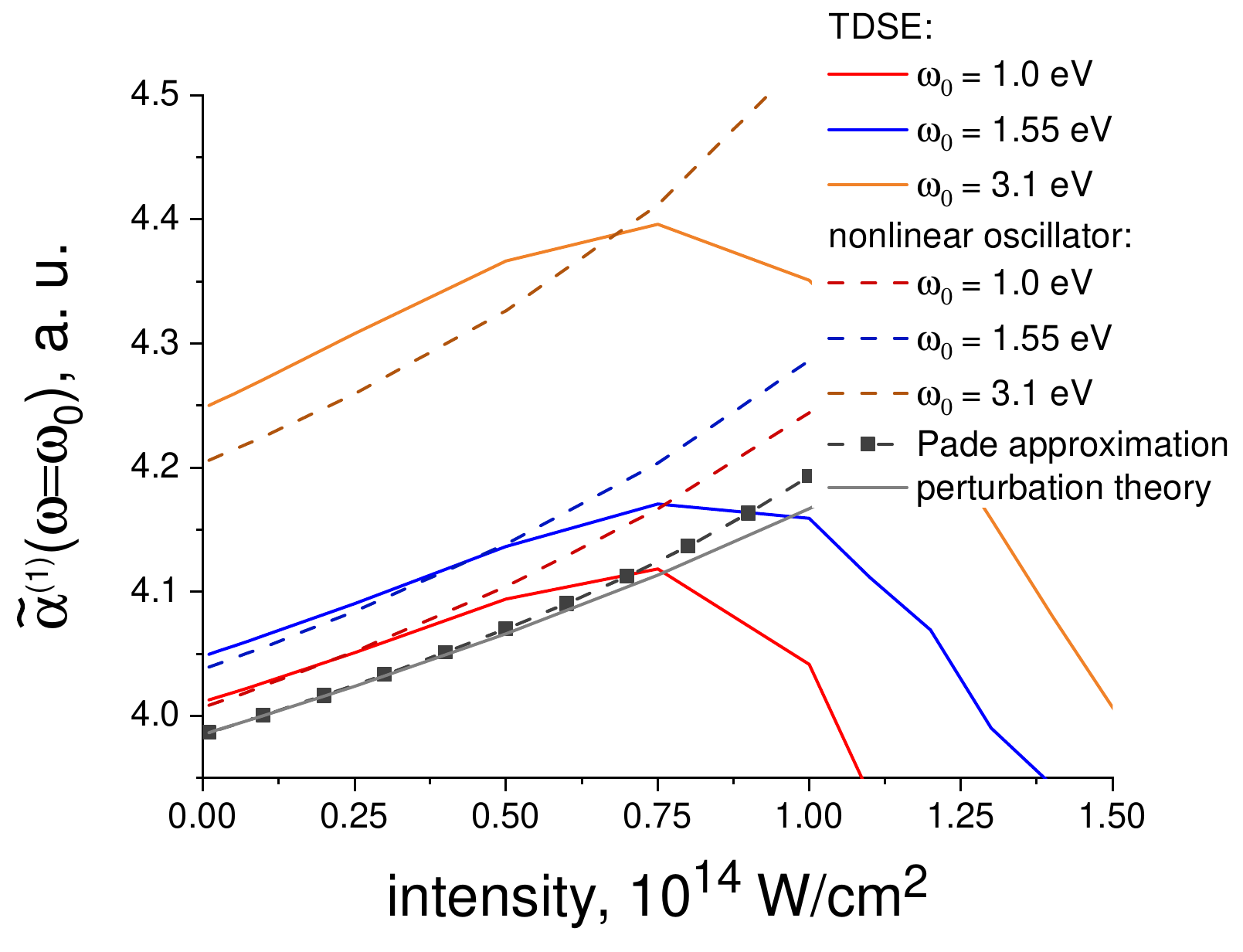}
    
    b)
    
    \includegraphics[width=0.8 \linewidth]{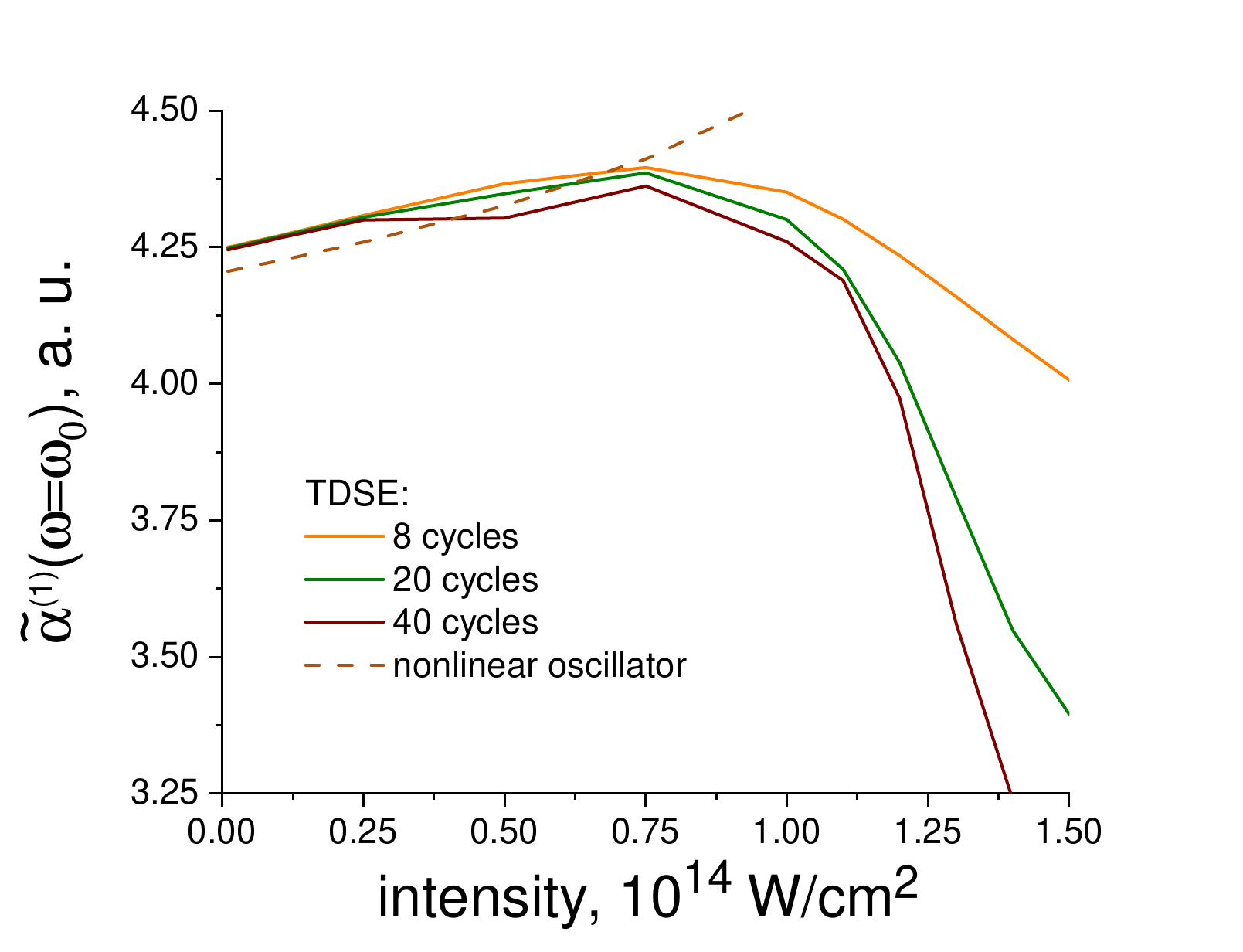}
    \caption{The dependence of the effective polarizability of the first order $\tilde\alpha^{(1)}(\omega_0)$ on the intensity for different frequencies (a) and for different pulse durations for $\omega_0=3.1~eV$ (b). The results of the TDSE numerical integration (solid curves), of the quasi-static model using Pad\'e approximation (black dashed curve  with symbols), of the nonlinear oscillator model (dashed curves without symbols), of the perturbation approach (solid grey curve).}
    \label{fig:chi1}
\end{figure}

\begin{figure}
    \centering
    \includegraphics[width=0.8 \linewidth]{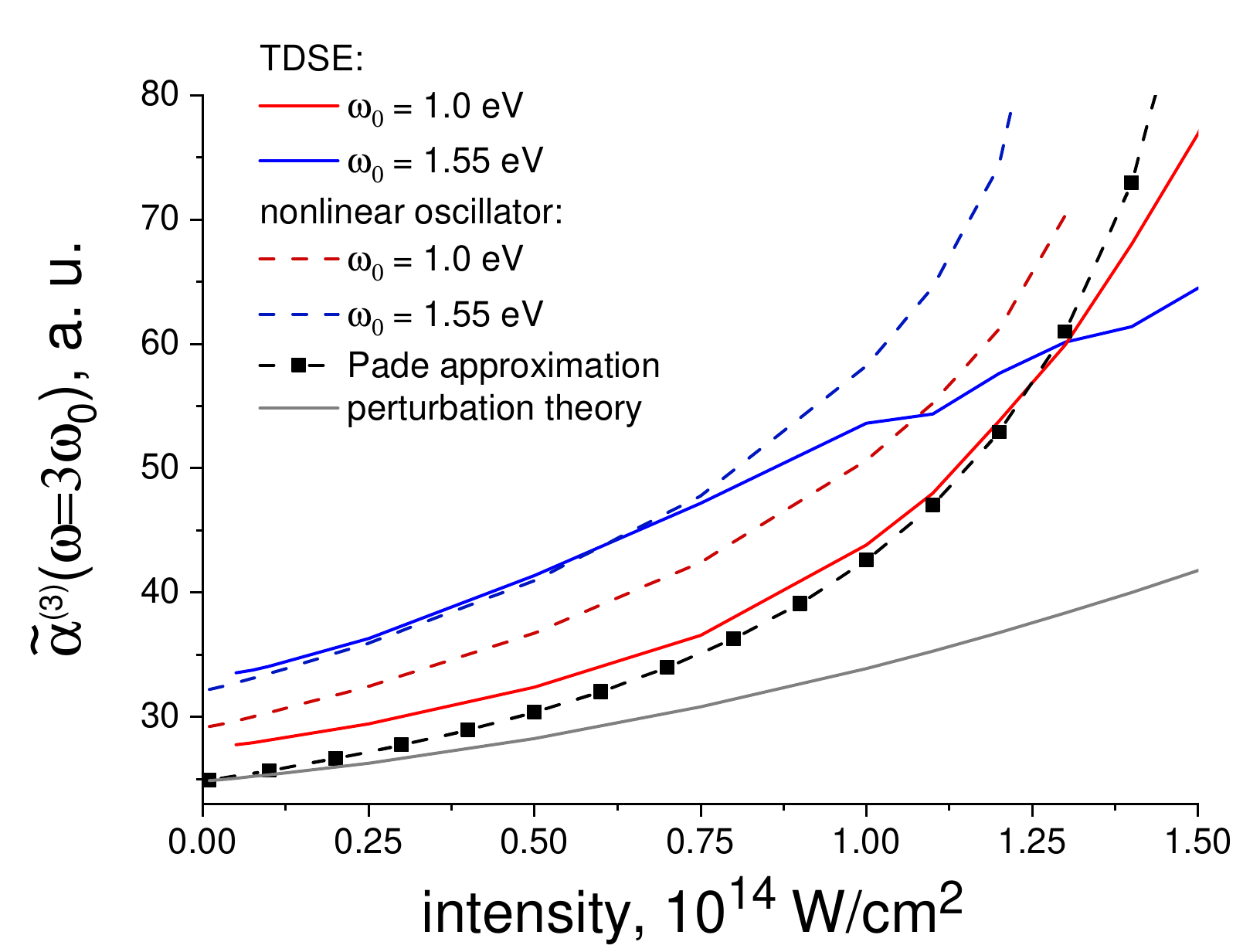}
    \caption{The same as Fig.~\ref{fig:chi1}a) for the third-order effective polarizability $\tilde\alpha^{(3)}(3\omega_0)$.}
    \label{fig:chi3}
\end{figure}

\begin{figure}
    \centering
    \includegraphics[width=0.8\linewidth]{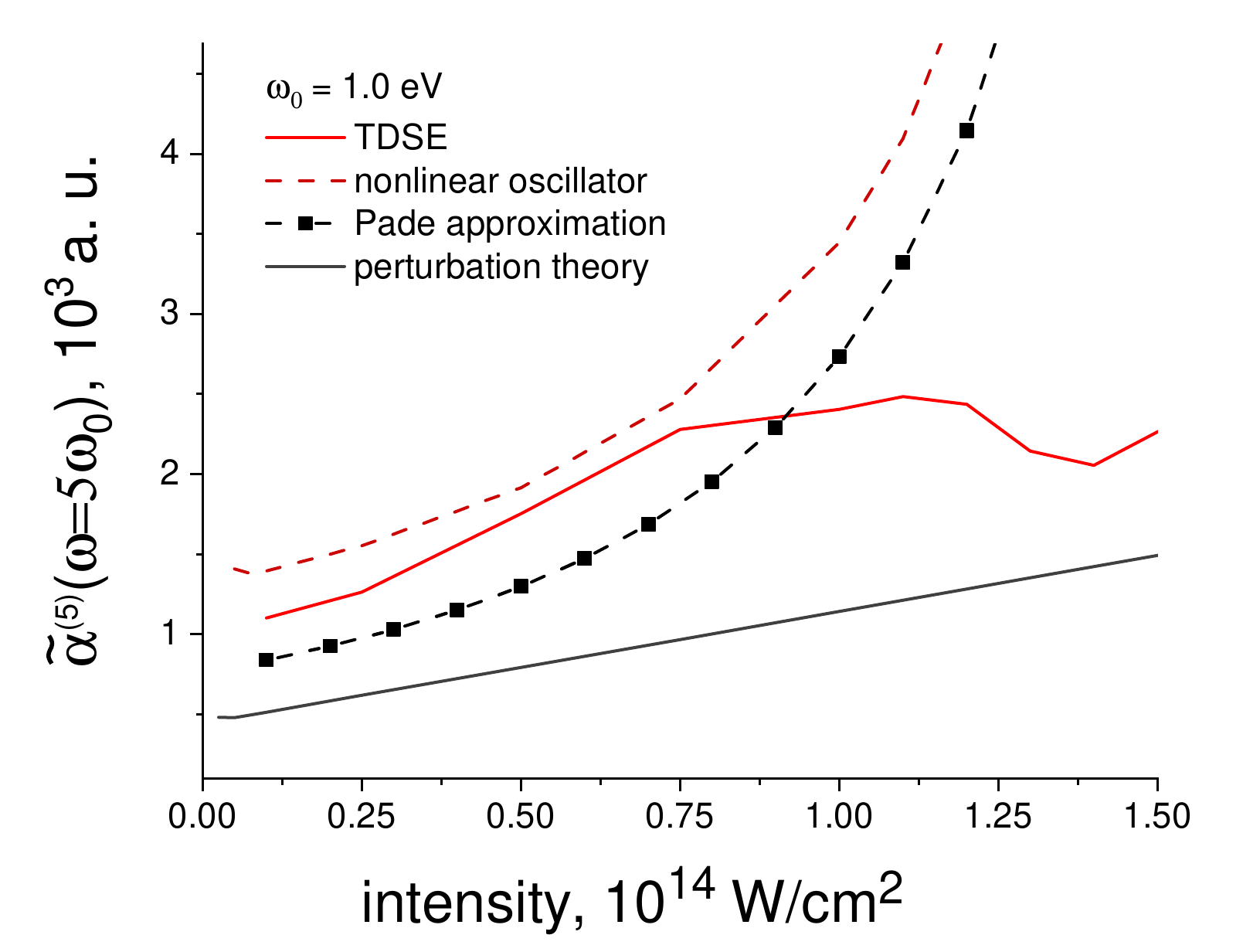}
    \caption{The same as Fig.~\ref{fig:chi1}a) for the fifth-order effective polarizability $\tilde\alpha^{(5)}(5\omega_0)$.}
    \label{fig:chi5}
\end{figure}

In the following figures we present the effective polarizabilities of the first (Fig.~\ref{fig:chi1}), third (Fig.~\ref{fig:chi3}), and fifth (Fig.~\ref{fig:chi5}) orders. 

In Fig.~\ref{fig:chi1} we see that for low intensity and low frequency (1 eV) the TDSE result agrees with those of the quasi-static and nonlinear oscillator. For higher frequencies the quasi-static model's applicability is worse, but the nonlinear oscillator model is still applicable, thus it adequately describes the dispersion. However, the TDSE result grows approximately linearly with intensity up to approximately 0.7~10$^{14}$ and then suddenly decreases. It does not show the {\it nonlinear} growth predicted by the models.

The linear growth of the refractive index (defined by the first order effective polarization) with intensity is known as the Kerr effect. When the intensity increases, the photoelectrons start contributing to the refractive index. This contribution is negative, so it leads to a decrease of the refractive index at higher intensities. The modulus of this contribution is higher for lower frequencies, so for lower frequencies this decrease is observed at lower intensities, see Fig.~\ref{fig:chi1}a. The compensation of the Kerr effect by photoelectrons underlies the the so-called standard model of the laser pulse filamentation in gases~\cite{Brodeur_1997, Mlejnek_1998, Couairon_2000, Berge_2007, Couairon_2007} and describes this process well~\cite{Kosareva_2011, Chin_2012, Kandidov_2013, Vrublevskaya_2023, Nikolaeva_2026}, at least in the IR domain.  

In Fig.~\ref{fig:chi1}a we see that for the UV field the effective polarization decreases with the intensity for high intensities, but this decrease cannot be completely attributed to the photoelectron contribution. Namely, this contribution increases linearly with the pulse duration; however, we see that the behaviour of the polarizability is similar for the  20- and 40-cycle pulses. Other mechanisms defining the decrease of the nonlinear refractive index with intensity are studied in Refs.~\cite{Mechain_2004, Loriot_2009, Bejot_2010, Ettoumi_2010, Volkova_2012, volkova_nonlinear_2013, Smetanin_2016}. Our detailed calculations in this direction will be published elsewhere~\cite{Strelkov_UV_filamentation}. In general, the refractive index in UV fields of moderate intensity can be affected by multiphoton resonances. For instance, in our case the energy of three photons of 3.1 eV field is comparable with the first excitation energy (11.6 eV), so the non-resonant approach can hardly be applicable for describing the nonlinear refractive index defined by the third-order nonlinearity.  

\begin{figure}
    \centering
    \includegraphics[width=0.8 \linewidth]{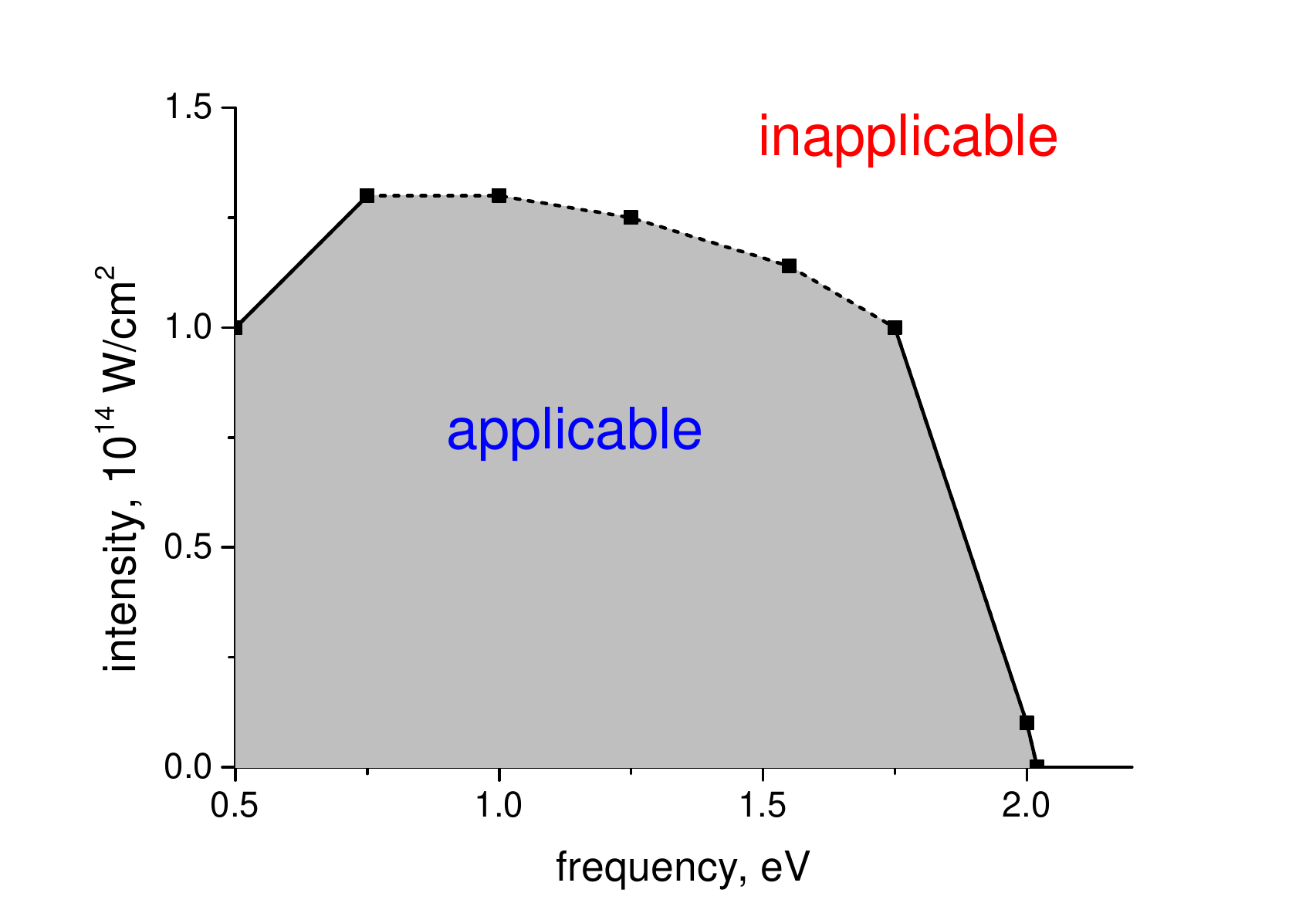}
    \caption{The applicability region of the nonlinear oscillator model for the third-order effective polarizability $\tilde\alpha^{(3)}(3\omega_0)$ calculation. The dashed line shows the region where the applicability is limited by the stability of the nonlinear oscillator behaviour (see text for more details), and the solid line shows the region where the applicability is limited by weak agreement with the TDSE results}
    \label{applicability}
\end{figure}

Thus, the non-perturbative increase of the effective first-order polarizability predicted by our models is not observed in the TDSE calculation. The reason is that for the intensities when this increase might take place, the photoelectrons affect the polarizability for the IR fields and resonant processes affect it for UV fields, whereas our models take into account neither photoionization nor resonances. For the third- and fifth-order nonlinearities the agreement for the UV field is even worse, so these results are not shown in Figs.~\ref{fig:chi3} and~\ref{fig:chi5}. However, for IR field the non-perturbative increase of these polarizabilities predicted by the models is confirmed by the TDSE calculations, see Figs.~\ref{fig:chi3} and~\ref{fig:chi5}. Note that the TDSE calculation in Ref.~\cite{ABecker2015} also shows an increase of the third- and fifth-order nonlinearity for similar intensities of the IR field. 

 Figure~\ref{applicability} shows the ranges of the laser frequencies and intensities where the nonlinear oscillator model can be used to calculate $\tilde\alpha^{(3)}(3\omega_0)$. For high frequencies (more than approximately 2 eV) the result of the model does not agree with the TDSE result (we assume that the disagreement takes place if the difference is more than 20$\%$). This limitation can be attributed to wrong description of the resonant contribution in the model; this contribution becomes significant when the laser frequency multiplied by the order of the process under consideration (in this case 3) becomes comparable to the transition frequency. For lower frequencies the nonlinear oscillator model agrees with the TDSE results up to a certain intensity ($1 \div 1.3$ x 10$^{14}$~W/cm$^2$), for which oscillation at the natural frequency suddenly starts and grows exponentially. Finally, for low frequencies (approximately 0.5 eV and lower) the pulse duration (which was eight periods in our calculations) becomes so long that the photoionization contributes to the TDSE result, so the results of the oscillator model and the TDSE do not agree even for intensities when the oscillator model is still stable.    

\begin{figure}
    \centering
    \includegraphics[width=0.8\linewidth]{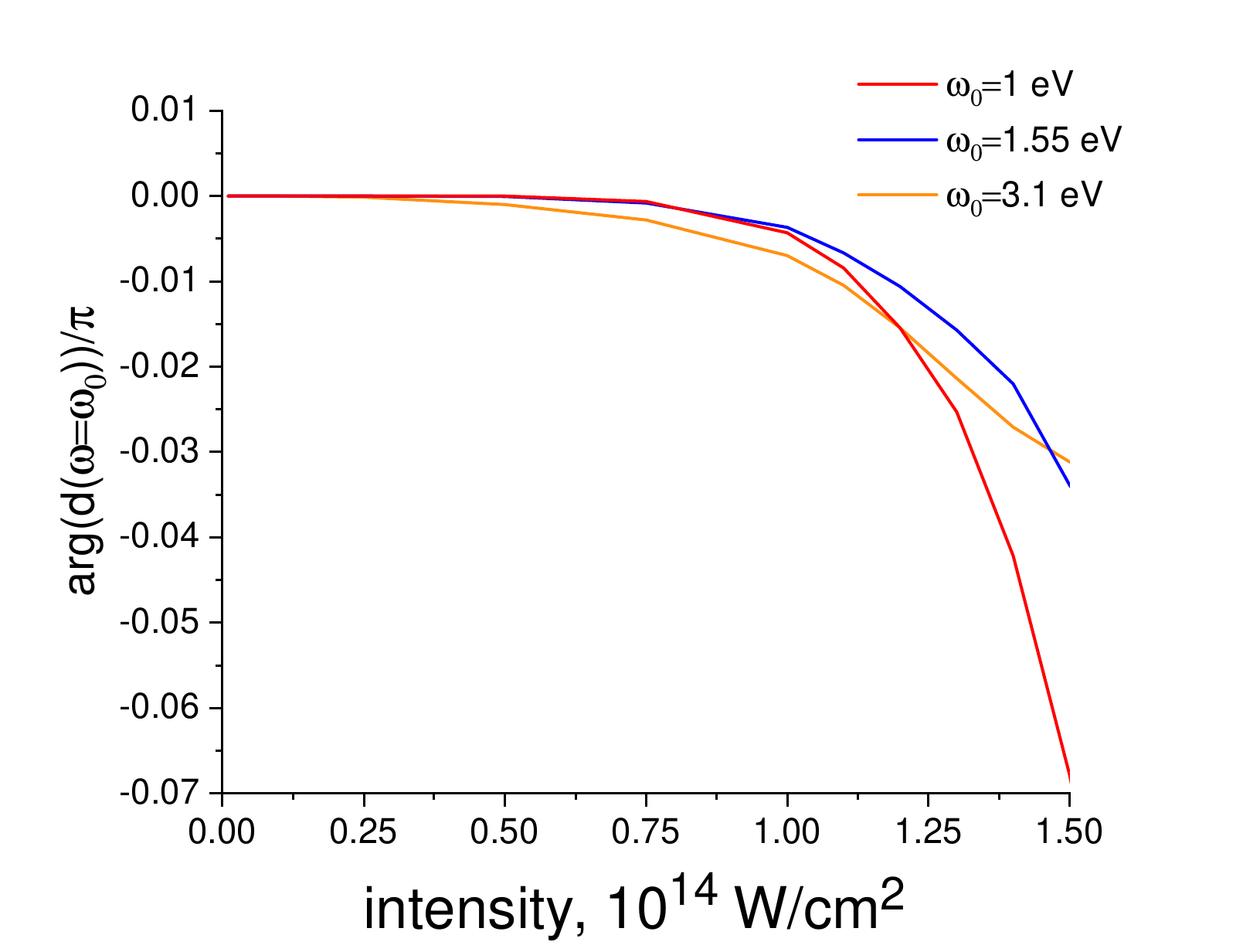}
    \caption{Phase of the first harmonic vs. laser intensity for different frequencies found using TDSE numerical solution.}
    \label{fig:argchi1}
\end{figure}

\begin{figure}
    \centering
    \includegraphics[width=0.8\linewidth]{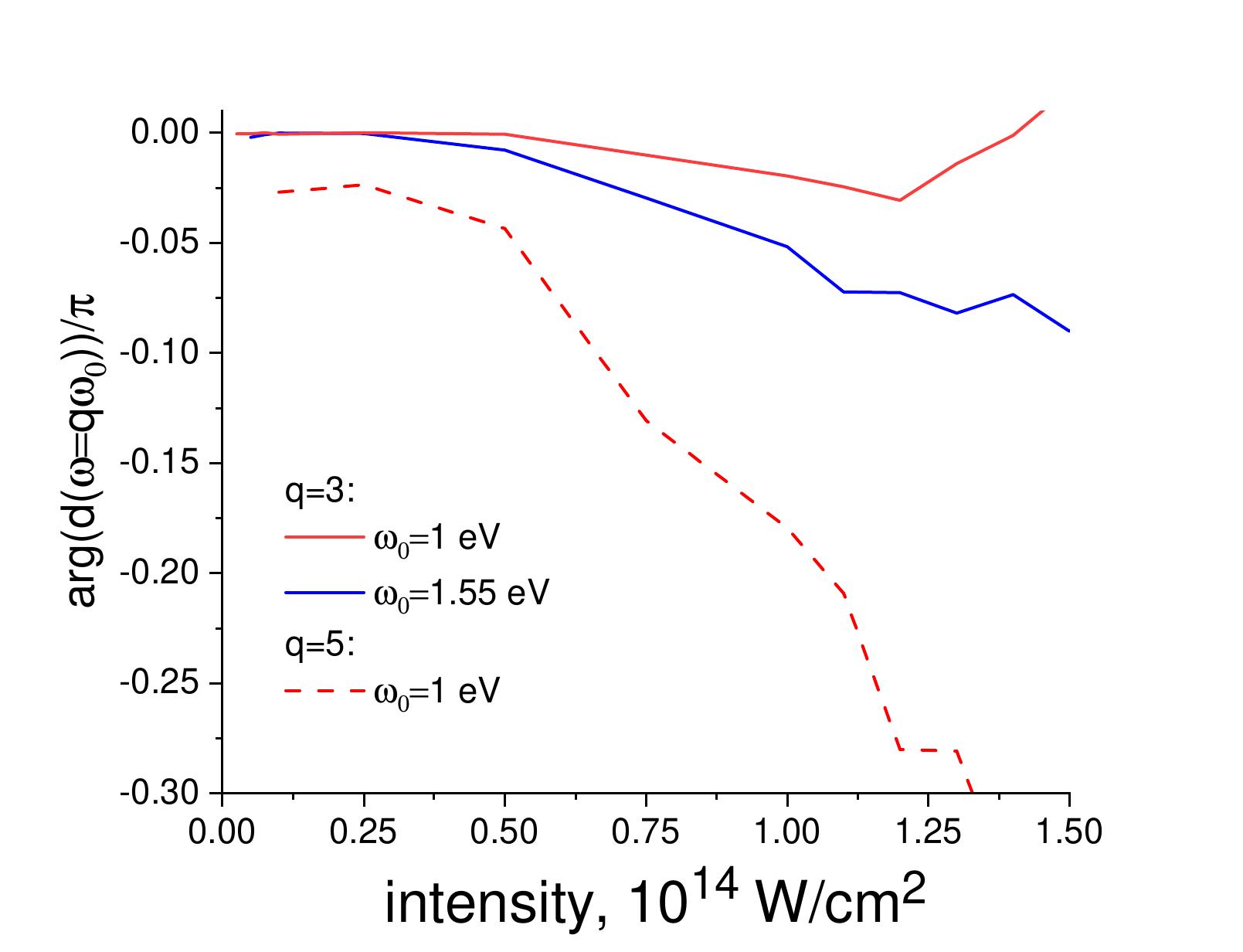}
    \caption{ The same as Fig.~\ref{fig:argchi1} for the third and fifth harmonic.}
    \label{fig:argchi35}
\end{figure}

Figs.~\ref{fig:argchi1} and~\ref{fig:argchi35} show the dependence of the harmonic phase $\arg (d(\omega=q\omega_0))$ on the laser intensity which was found using numerical TDSE solution for different laser frequencies.
For low intensities the atomic response adiabatically follows the electric field, so the harmonic phases are zero. For higher intensities the response is slightly delayed with respect to the driving field (in agreement with the findings of Refs.~\cite{Vrublevskaya_2023, Shipilo_2025}).%, see the inset in Fig.~\ref{fig:argchi1}.
This delay leads to the non-zero harmonic phase for high intensities in Figs.~\ref{fig:argchi1} and~\ref{fig:argchi35}. 

The non-zero harmonic phase is found only in the numerical TDSE solution. The quasi-static model supposes instantaneous response to the field and so naturally predicts zero harmonic phases. Counter-intuitively, the nonlinear oscillator model also does not describe the harmonic phase adequately. Namely, under the steady-state conditions (assuming, in particular, that the ionization is negligible) the oscillator does not gain energy from the field, so the field does not do any work, and the phase of the response at the fundamental frequency is zero. Our analytical study of the nonlinear oscillator (see Appendix C) shows that that in this case the harmonic phase is also zero. Moreover, in the numerical simulation of the nonlinear oscillator motion the non-zero phases are found only in a narrow range of intensities close to the ionization threshold.

Below we consider the nonlinear response to a two-color field:
\begin{equation}
\begin{array}{cc}
    E(t) = E_0 g(t) [ \cos (\omega_0(t-t_{full}/2))+\\ \nu \cos (2\omega_0(t-t_{full}/2)+\Delta \phi) ], 
    \label{two-color}
\end{array}
\end{equation}
where $E_0$ is the amplitude of the fundamental (first harmonic) field, $\nu$ is the amplitude ratio between the second harmonic and the fundamental, %$g(t)$ is a slowly varying envelope function that defines the pulse shape, $t_{full}$ denotes the full duration of the pulse, 
$\Delta \phi$ is the phase difference. We study the third-order nonlinear process of optical rectification or generation of a quasi-static field. This process is used, in particular, for the generation of terahertz fields~\cite{Vvedenskii_2014,Kostin2016}.

The effective third-order nonlinear polarizability associated with this process is:
\begin{equation}
   \tilde\alpha^{(3)}(\omega=2\omega_0-\omega_0-\omega_0)=\frac{\bar d(\omega=2\omega_0-\omega_0-\omega_0)}{\bar E(2\omega_0)  \left( \bar E(\omega_0) \right) ^2}.
\end{equation}
Note that this effective third-order nonlinear polarizability depends on phase difference $\Delta \phi$,   in agreement with the experimental findings of Ref.~\cite{Zhang_2025}. 

\begin{figure}
    \centering
    a)
    
    \includegraphics[width=0.8\linewidth]{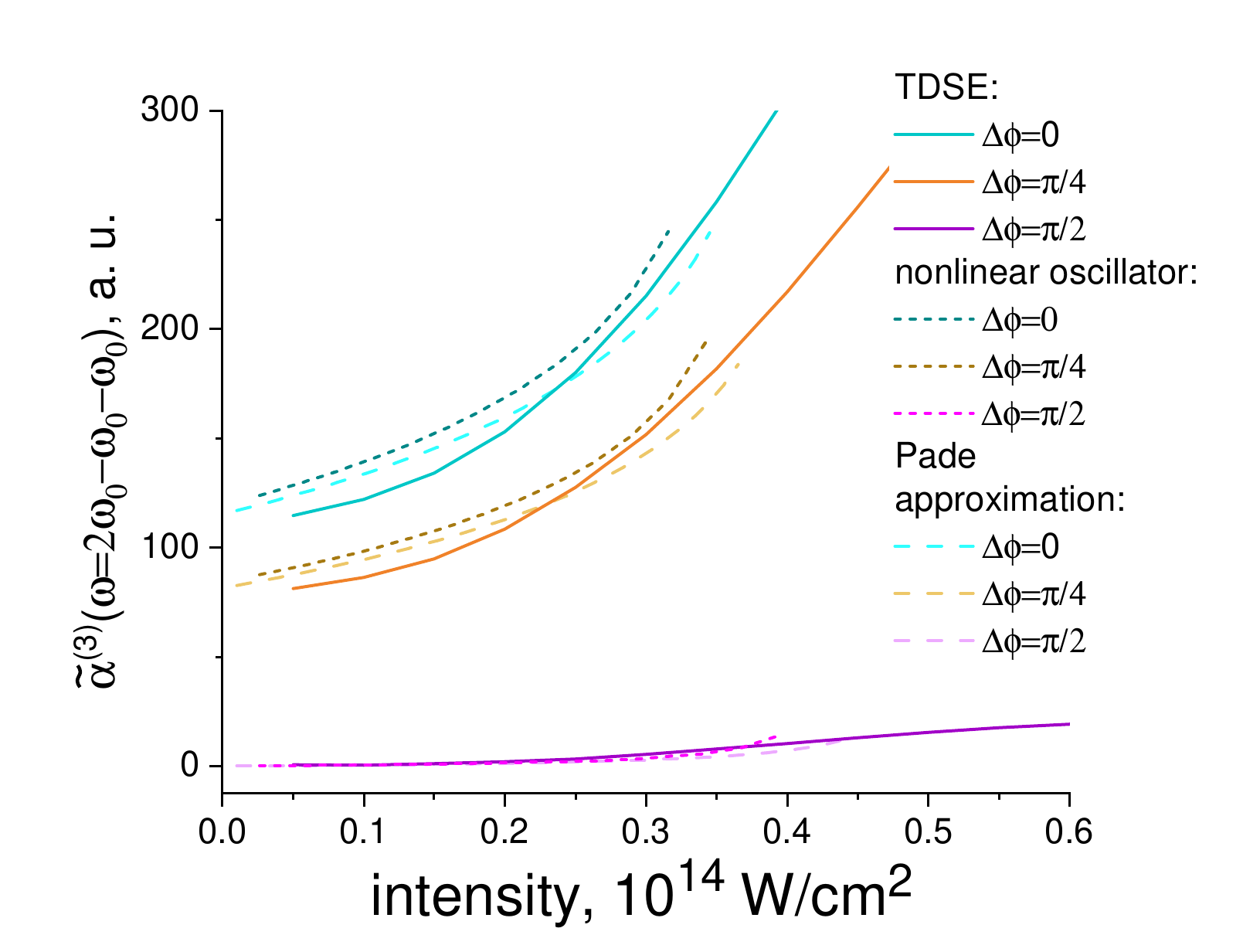}
    
 b)

    \includegraphics[width=0.8\linewidth]{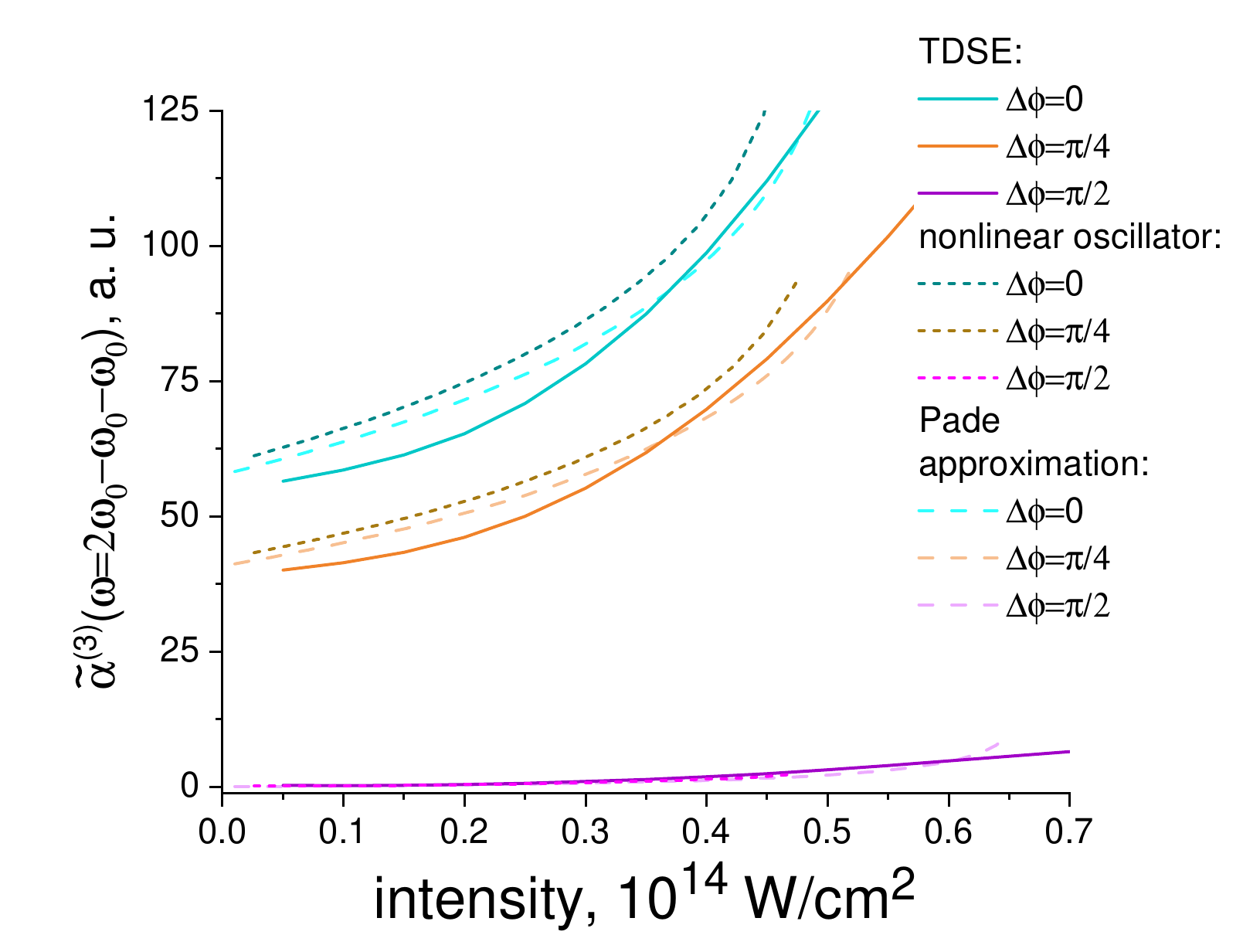}
    
    \caption{The effective polarizability $\tilde\alpha^{(3)}(\omega=2\omega_0-\omega_0-\omega_0)$ describing optical rectification process as function of the fundamental intensity~($\propto E_0^2$) for various phase differences $\Delta \phi$ for two-color generating field given by Eq.~(\ref{two-color}). The fundamental frequency is $\omega_0=1$~eV, the  ratio of the generating fields' amplitudes are $\nu=1$ (a) and $\nu=1/\sqrt{2}$ (b).}
    \label{fig:two-color}
\end{figure}

Fig.~\ref{fig:two-color} presents the polarizability found within the TDSE numerical solution, the quasi-static model, and the nonlinear oscillator model. Graphs a) and b) show the results for different ratios of the generating fields' amplitudes.  We see that polarizability rapidly grows with the laser intensity and strongly depends on the phase difference. Both features are well reproduced by the quasi-static model and the nonlinear oscillator model. Note that the intensity shown along the horizontal axis is that of the first harmonic only. The applicability of our models is limited by the maximal strength of {\it total} field, achieved already at moderate intensities of the fundamental.

\section*{Discussion and Conclusions}

We solve numerically three-dimensional non-stationary Schr\"odinger equation for a model argon atom in a laser field and find the atomic response for different field frequencies. In particular, we find the atomic dipole moment in a quasi-static electric field. For weak fields, the response as a function of the field is described by the Taylor expansion, but for the intensities higher than  a certain value (about $0.6 \,  \,  10^{14}$~W/cm$^2$ for argon) the accuracy of the approximation with the Taylor (and other polynomial) series is unsatisfactory, regardless of the order of non-linearity. This defines a limit of the perturbative description of the atomic response in the quasi-static field. We suggest Pad\'e approximation of the atomic dipole moment which agrees with the TDSE numerical calculation result in a quasi-static field, both for low and high intensities approximately up to  half of the barrier-suppression intensity (in particular, for argon the approximation is applicable up to $1.4 \,  \,   10^{14}$ W/cm$^2$); for higher intensities the photoionization-induced processes provide the main contribution to the nonlinearity and our approximation becomes inapplicable. We use this expression to calculate effective nonlinear polarizabilities in the quasi-static limit. Moreover, we use it for calculating the restoring force, suggesting the nonlinear oscillator model to describe the dispersion of the nonlinearities beyond the perturbation approach. Our model fails to describe the nonlinear refractive index in the nonperturbative domain, in terms of both amplitude and phase, because it doesn't take into account the contribution of either the photoelectrons (dominating for the IR fundamental) or resonances (dominating for the UV fundamental). In contrast, the third and fifth harmonic generation in the IR field, as well as optical rectification in a two-color field, are well described by our model, including the non-perturbative growth of the efficiency with the laser intensity. For these processes as well,  the applicability of the model for high intensities and frequencies is limited by photoionization and resonant contribution. 

 The increase of the third and fifth-order nonlinearity with the laser intensity can be understood within our model as follows. The bound electron oscillating in the intense field finds itself far from the origin (namely, at the distances comparable to the initial size of the bound wave-packet) where the Coulomb attraction is reduced. This leads to an increase of the nonlinear polarizability.

 Hence the microscopic response calculated via TDSE solution can be directly used in propagation simulation~\cite{KhokhlovaStrelkov, Wang_2023, Birulia2024, Vabek2026}, the dimensionality reduction (in TDSE or/and macroscopic equations) is necessary to make propagation simulations feasible. The model suggested in this paper can be easily integrated with full 3D macroscopic propagation solvers. It can hardly describe the self-action better than the perturbation approach does. However, it can be useful to study macroscopic efficiency of the third- and fifth- order processes (third and fifth order harmonic generation, optical rectification, four-wave mixing, etc.). In particular, the non-perturbative growth of the microscopic nonlinearity with the fundamental intensity predicted by our model can lead to a practically important enhancement of the macroscopic signal compared to the perturbation theory prediction.

\section*{Acknowledgment}
This study was funded by RSF through Grant No. 24-12-00461.

\section*{Appendix}
\subsection{Details of the TDSE numerical solution}

In the Hamiltonian given by Eq.~(\ref{eq: hamiltonian})  for argon we use the model potential $U(r)$ similar to the one suggested in~\cite{Strelkov_2005}:  
\begin{equation}
    U(r) = - \frac{1 + \alpha e^{-\sqrt{r^2 + \beta^2}}}{\sqrt{r^2+ r_0^2}},
    \label{eq: atomic_potential}
\end{equation}
with parameters: $\alpha=5.94,~\beta=0.2, ~r_0=2.105$. This potential reproduces the ionization energy of the argon atom. Moreover, the energies of several lowest excited states are close to these energies in the actual atom. 

 For neon we use the model potential 
\begin{equation}
    U(r) = - \frac{1 + e^{-r/R_0}}{\sqrt{r^2+r_0^2}},
    \label{eq: Ne_potential}
\end{equation}
with $R_0=2.7$ and $r_0=0.5$. 

 For xenon we use the soft Coulomb potential:
\begin{equation}
    U(r) = - \frac{1}{\sqrt{r^2+r_0^2}},
    \label{eq: Xe_potential}
\end{equation}
with $r_0=0.1$.

 The potentials given by Eqs.~(\ref{eq: Ne_potential}) and (\ref{eq: Xe_potential}) reproduce the ionization energies of the neon and xenon atoms, respectively. 

The TDSE~(\ref{eq: tdse}) is solved numerically by the Crank-Nicolson method, see more details in~\cite{Strelkov_2006}. The size of the numerical box is $[-z_b, z_b]$ in the longitudinal direction and $[0, \rho_b]$ in the radial direction. The spatial step is 0.2 a.u. in both directions, and the temporal step is 0.04 a.u. 

The imaginary absorbing potential $-iW(\rho, z)$ in Eq.~(\ref{eq: hamiltonian}) is used to prevent unphysical reflections of the wave packet from the boundaries of the numerical box. It is written as follows:
\begin{equation}
    W(\rho, z) = W_z(z) + W_\rho(\rho),
    \label{eq: abs_total}
\end{equation}
where
\begin{equation}
    W_z(z) =
\left\{
\begin{array}{ll}
        \frac{|z| - z_{\text{abs}}}{z_b - z_{\text{abs}}} W_0, ~\text{if}~|z| > z_{\text{abs}}, \\
        0, ~\text{if}~|z| < z_{\text{abs}},
\end{array}
\right.
    \label{eq: abs_z}
\end{equation}
and
\begin{equation}
    W_\rho(\rho) =
\left\{
\begin{array}{ll}
        \frac{\rho - \rho_{\text{abs}}}{\rho_b - \rho_{\text{abs}}} W_0, ~\text{if}~\rho > \rho_{\text{abs}}, \\
        0, ~\text{if}~\rho < \rho_{\text{abs}},
\end{array}
\right.
    \label{eq: abs_r}
\end{equation}
$W_0 = 1$ a.u. is the maximum of the absorbing potential, the width of the absorbing layer is $z_b-z_{\text{abs}} = \rho_b - \rho_{\text{abs}}= 6$~a.u.

\begin{figure}%[h]
    \centering
    \includegraphics[width=0.8\linewidth]{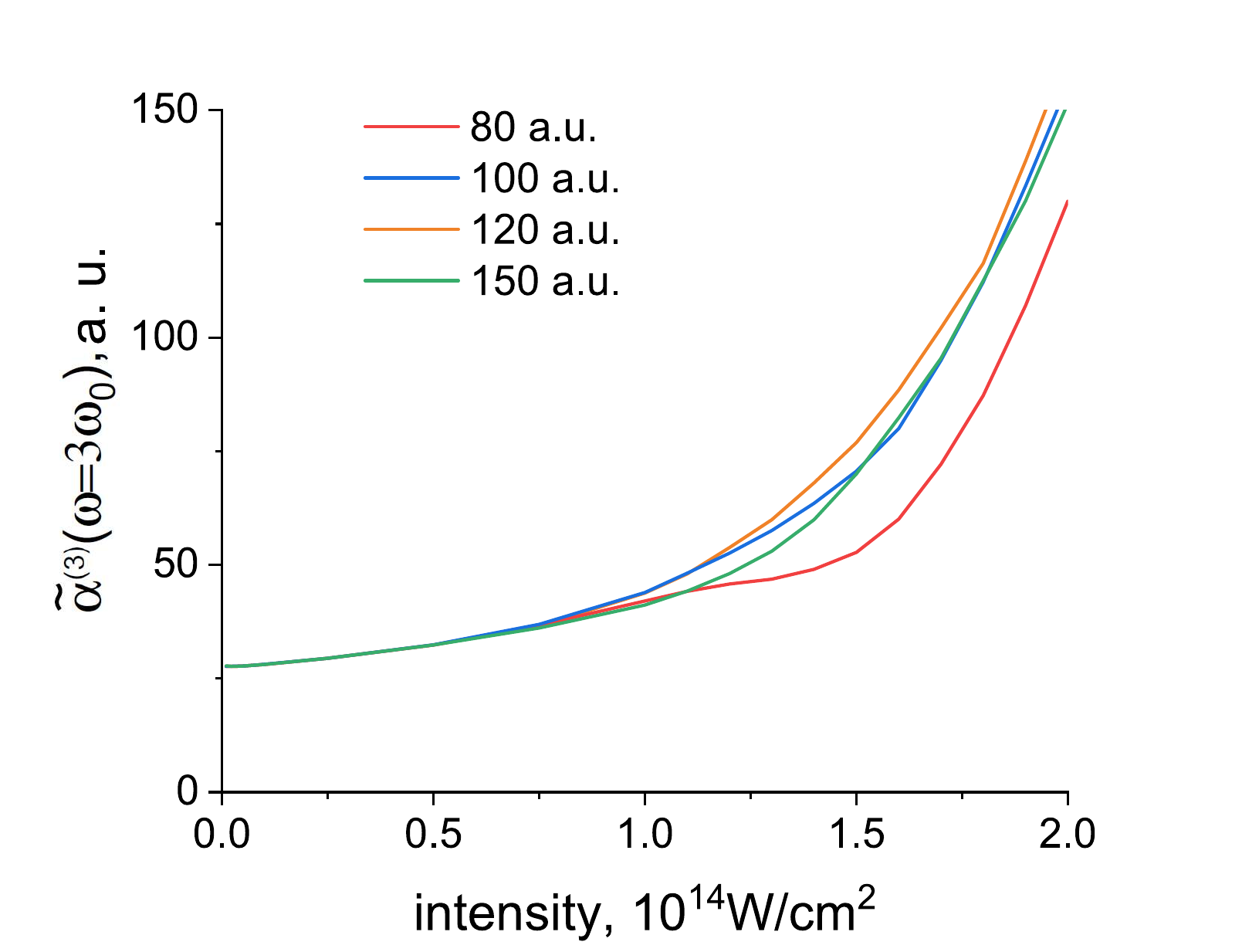}
    \caption{ The effective polarizability $\tilde \alpha^{(3)}(3 \omega_0)$ calculated for $\omega_0 = 1$ eV via the TDSE using different sizes of the numerical box $z_b$ in the longitudinal direction, shown in the graph (the box is $[-z_b, z_b]$ in the longitudinal direction and $[0, 60 \  a.u.]$ in the radial direction).}
    \label{num_box}
\end{figure}

In the TDSE calculation of the quasi-static response (Subsection~\ref{num_q-s}) we use a confined numerical box to find only the response of the bound electron, so we use $z_b=20$ a.u. In the other calculations we are interested in the complete response, thus including the contributions from both the bound state and the free wave-packets. The electron wave-packet after the photoionization oscillates in the laser field with the amplitude $z_{osc}=E_0/\omega_0^2$ and spreads. To check if the used numerical box is large enough to provide adequate result we make calculations for different longitudinal sizes of the numerical box, see Fig.~\ref{num_box}. For the used frequency and for the highest used intensity we have $z_{osc}=55$ a.u. In the figure one can see that the calculated results almost do not depend on $z_b$ as soon as it exceeds 100 a.u., thus when $z_b>2 z_{osc}$. So in our calculations for laser frequencies 1.55 eV and higher we used $z_b=60$~a.u. and for lower  frequencies we used $z_b=2 z_{osc}$, where  $z_{osc}$ is found for the highest laser intensity under consideration. Finally,  $\rho_b=60$~a.u. is used in all calculations. This is sufficient for describing the free wave-packet motion during several field cycles, taking into account that the transverse spread of the wave-packet is given by the uncertainty of the photoelectron's transverse momentum: $\Delta p_{\perp} = \sqrt{2 I_p E}$ (see~\cite{Perelomov}), where $I_p$ is the ionization potential.

\subsection{Chebyshev approximation}
The Chebyshev polynomial of degree $N$ is given by:
\begin{equation}
    T_N(x)=\cos(N~\arccos x).
\end{equation}
%Chebyshev polynomials form an orthogonal set in the interval [-1, 1] with the weight function $1/\sqrt{1-x^2}$.
The polynomial $T_N(x)$ has $N$ zeros in the interval [-1, 1] and they are written in the following form:
\begin{equation}
    x_j^{(N)}=\cos \left( \frac{\pi (j+1/2)}{N} \right), j=0, 1, \dots, N-1.
    \label{zeros}
\end{equation}
Any continuous function $f(x)$ defined on [-1, 1] can be approximated by a finite Chebyshev series~\cite{press2007numerical}:
\begin{equation}
    f(x) \approx \sum_{m=0}^{N-1} \beta_m T_m(x)-\frac{1}{2}\beta_0, 
    \label{eq:chebyshev_series}
\end{equation}
where $N-1$ is the maximum degree of the Chebyshev polynomial, the coefficients $\beta_m$ are found as:
\begin{equation}
    \beta_m=\frac{2}{N} \sum_{j=1}^{N-1}  f(x_j^{(N)}) T_m(x_j^{(N)}),
    \label{beta}
\end{equation}
and $x_j^{(N)}$ are given by Eq.~(\ref{zeros}).

Similarly to Eq.~(\ref{eq:chebyshev_series}) we decompose the dipole response to the quasi-static field $d(E)$ into Chebyshev series up to the seventh order in the interval $E=[-E_{\text{max}}, E_{\text{max}}]$ as follows:
\begin{equation}
\begin{array}{cc}
     &d(E)\approx\beta_1T_1(E/E_{\text{max}})+\beta_3T_3(E/E_{\text{max}})+\\
     & +\beta_5 T_5(E/E_{\text{max}})+\beta_7 T_7(E/E_{\text{max}}) \equiv \\
     &  \kappa^{(1)} E+\kappa^{(3)}E^3+\kappa^{(5)} E^5+\kappa^{(7)} E^7.
\end{array}
\label{eq: chebyshev}
\end{equation}
Substituting the explicit form of the Chebyshev polynomials we find from the latter Eq.:
\begin{equation}
\begin{array}{cc}
    &\kappa^{(1)}=(\beta_1-3\beta_3+5\beta_5-7\beta_7)/E_{\text{max}}, \\
    &\kappa^{(3)}=(4\beta_3-20\beta_5+56 \beta_7)/E^3_{\text{max}}, \\
&\kappa^{(5)}=(16\beta_5-112 \beta_7)/E^5 _{\text{max}}, \\
    &\kappa^{(7)}=64 \beta_7/E^7 _{\text{max}}.
\end{array}
\label{kappa_beta}
\end{equation}

From Eq.~(\ref{beta}) we find coefficients $\beta_m$ using numerically calculated $d(E)$ and $E_{\text{max}}=0.033$ a.u. From Eq.~(\ref{kappa_beta}) we find the following values of $\kappa^{(m)}$:
\begin{equation}
\begin{array}{cc}
    &\kappa^{(1)}\approx 3.99, \\
    &\kappa^{(3)}\approx 96, \\
&\kappa^{(5)} \approx5.8~10^3, \\
    &\kappa^{(7)}\approx 2~10^6 .
\end{array}
\end{equation}

\subsection{Duffing oscillator model}
We expand the restoring force in Eq.~(\ref{eq:oscillator}) up to the third order:
\begin{equation}
    F(d)= -\Omega_0^2 d (1+\gamma d^2),
\end{equation}
where $\Omega_0^2=1/\kappa^{(1)}$,  $\gamma=-\kappa^{(3)}/(\kappa^{(1)})^3$. Then Eq.~(\ref{eq:oscillator}) for the monochromatic driving force is written as:
\begin{equation}
    \ddot{d}+\Omega_0^2 d (1+\gamma d^2) = E_0\cos(\omega_0 t), 
    \label{eq: oscillator_duffing}
 \end{equation}
Introducing
\begin{equation}
\begin{array}{ll}
\Omega_0t=\tau \\
\Omega=\omega_{0}/\Omega_0 \\
\varepsilon=E_0/\Omega_0^2,
\end{array}    
\end{equation}
we write the latter Eq. as:
\begin{equation}
    \ddot{d}+d+\gamma d^3 = \varepsilon\cos(\Omega \tau). 
    \label{oscillator}
 \end{equation}
This equation describing an oscillator with cubic nonlinearity is called the Duffing equation\cite{duffing}.
 
Let us assume that the response contains only the first and third harmonics:
\begin{equation}
    d(\tau)=d_1\cos(\Omega \tau+\varphi_1)+d_3\cos(3\Omega \tau+\varphi_3), 
    \label{eq:sol_duffing}
\end{equation}
where $d_1,~d_3$ are the amplitudes of the first and third harmonics and $\varphi_1,~\varphi_3$ are the phases of these harmonics. 

Substituting (\ref{eq:sol_duffing}) into (\ref{eq: oscillator_duffing}), neglecting components at multiple excitation frequencies (except the third one) and collecting coefficients at the first and the third harmonic frequencies, we find:
{\small
\begin{equation}
\left\{
\begin{array}{ll}
    (1-\Omega^2)d_1+\frac{3d_1^3\gamma}{4}+\frac{3d_1 d_3^2\gamma}{2}+\frac{3d_1^2d_3\gamma}{4} \cos(\varphi_3-3\varphi_1) =\varepsilon\cos\varphi_1\\
    (1-9\Omega^2)d_3+\frac{3d_3^3\gamma}{4}+\frac{d_1^3\gamma}{4} \cos(\varphi_3-3\varphi_1)+\frac{3d_1^2 d_3\gamma}{2} \cos(\varphi_3-\varphi_1) =0\\
    \frac{d_1^3\gamma}{4} \sin(\varphi_3-3\varphi_1)+\frac{3d_1^2 d_3\gamma}{2} \sin(\varphi_3-\varphi_1)=0 \\
    \frac{3d_1^2 d_3\gamma}{4} \sin(3\varphi_1-\varphi_3)=\varepsilon\sin \varphi_1
\end{array}
\right.
\label{eq:z1_z3_0}
\end{equation}
}

From the latter two equations we obtain:
\begin{equation}
    \sin \varphi_1 = -\frac{9 d_1 d_3^2 \gamma}{2 \varepsilon} \sin(\varphi_1-\varphi_3).
\end{equation}
We see that one of the solutions of this equation is $\varphi_1=\varphi_3=0$.

\begin{figure}%[h]
    \centering
    \includegraphics[width=0.8\linewidth]{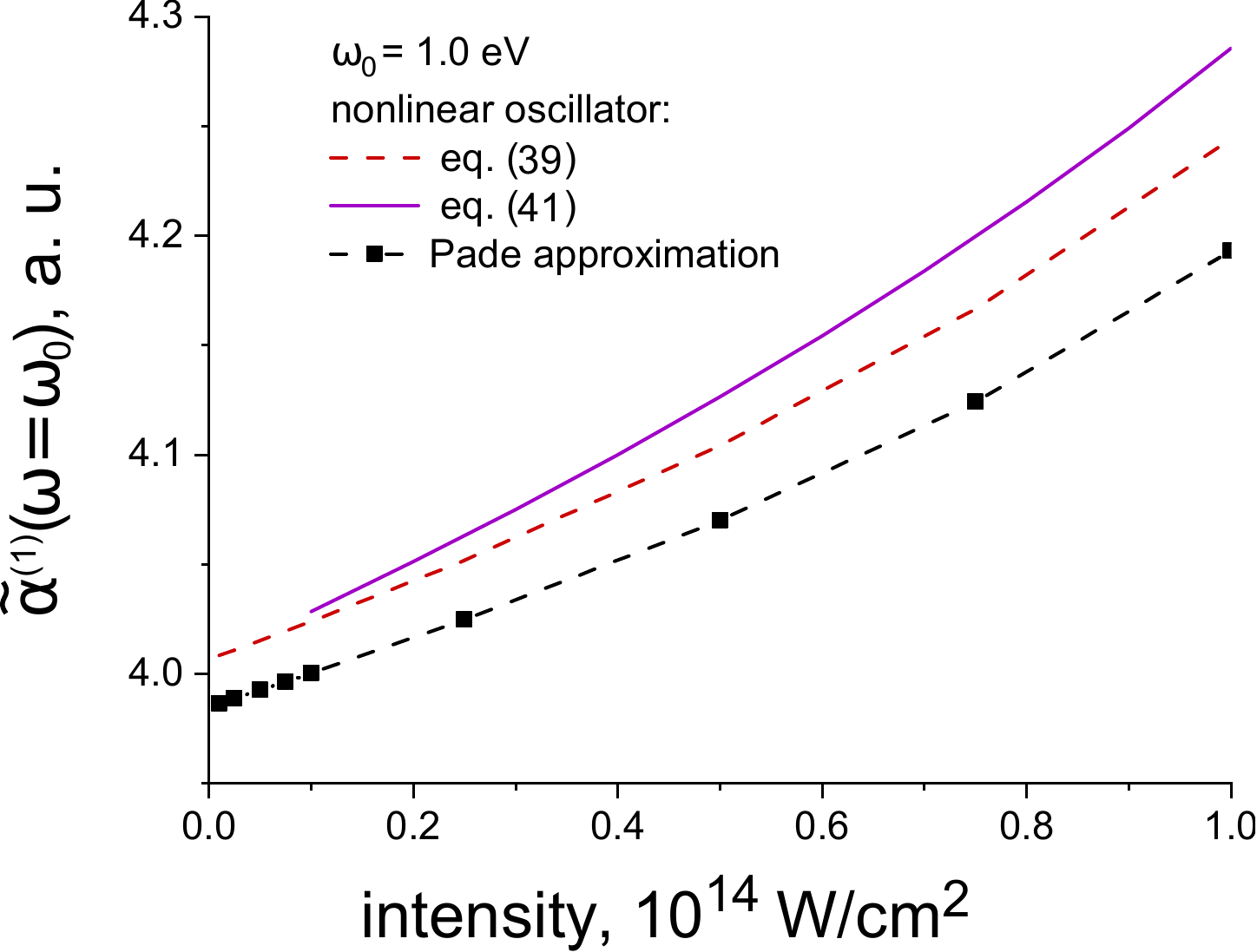}
    \caption{The effective polarizability $\tilde \alpha^{(1)}(\omega_0)$ {\it vs.} intensity calculated within the Duffing oscillator model for $\omega_0 = 1$ eV and within the model of the quasi-static response using Pad\'e approximation.  }
    \label{fig:z(t)}
\end{figure}

In this case, the system of equations is simplified and takes the form:
\begin{equation}
\left\{
\begin{array}{ll}
    (1-\Omega^2)d_1+\frac{3d_1^3\gamma}{4}+\frac{3d_1 d_3^2\gamma}{2}+\frac{3d_1^2d_3\gamma}{4}=\varepsilon\\
    (1-9\Omega^2)d_3+\frac{3d_3^3\gamma}{4}+\frac{d_1^3\gamma}{4} +\frac{3d_1^2 d_3\gamma}{2}  =0
\end{array}
\right.
\label{eq:z1_z3}
\end{equation}

We numerically solve Eq.~(\ref{eq:z1_z3_0}) and Eq.~(\ref{eq:z1_z3}) and find $d_1(\varepsilon)$ and $d_3(\varepsilon)$. Then we assume that these amplitudes vary adiabatically with the laser field envelope $g(t)$, so these slowly-varying amplitudes are  $d_1(g(t )\varepsilon)$ and $d_3(g(t)\varepsilon)$. Substituting these  amplitudes in Eq.~(\ref{eq:sol_duffing}) we obtain the response $d(t)$. Finally, from the Fourier transform of the response and of the field we calculate the polarizabilities. In Fig.~\ref{fig:z(t)} we compare $\tilde \alpha^{(1)}(\omega_0)$ obtained using solutions of Eq.~(\ref{eq:z1_z3_0}) and Eq.~(\ref{eq:z1_z3}). We can see that these results agree well.

%\subsection{ Convergence of the TDSE results with respect to the size of the numerical box}}

\section*{Data availability statement}
The data that support the findings of this article are openly available~\cite{data_nonperturbative}.

\bibliography{lit}% Produces the bibliography via BibTeX.

@article{3-step_C,
  title = {Plasma perspective on strong field multiphoton ionization},
  author = {Corkum, P. B.},
  journal = {Phys. Rev. Lett.},
  volume = {71},
  issue = {13},
  pages = {1994--1997},
  numpages = {0},
  year = {1993},
  month = {Sep},
  publisher = {American Physical Society},
  doi = {10.1103/PhysRevLett.71.1994},
  url = {https://link.aps.org/doi/10.1103/PhysRevLett.71.1994}
}

@book{Shen,
  title = {The principles of nonlinear optics},
  publisher = {Wiley-Interscience Publication},
  year = {1984},
  author = {Y. R. Shen},
  address = {New York},
  edition = {},
}

@article{Strelkov_2006,
	doi = {10.1088/0953-4075/39/3/011},
	url = {https://doi.org/10.1088%2F0953-4075%2F39%2F3%2F011},
	year = 2006,
	month = {jan},
	publisher = {{IOP} Publishing},
	volume = {39},
	number = {3},
	pages = {577},
	author = {V V Strelkov and A F Sterjantov and N Yu Shubin and V T Platonenko},
	title = {{XUV} generation with several-cycle laser pulse in barrier-suppression regime},
	journal = {J. Phys. B: At. Mol. Opt. Phys.},
}

@article{Lewenstein,
	author = {Lewenstein, M. and Balcou, {\relax Ph}. and Ivanov, M. {\relax Yu}. and L{'}Huillier, Anne and Corkum, P. B.},
	title = {Theory of high-harmonic generation by low-frequency laser fields},
	journal = {Phys. Rev. A},
	volume = {49},
	number = {3},
	pages = {2117--2132},
	year = {1994},
	month = {Mar},
	issn = {2469-9934},
	publisher = {American Physical Society},
	doi = {10.1103/PhysRevA.49.2117}
}

@article{KhokhlovaStrelkov,
	author = {Khokhlova, M. A. and Strelkov, V. V.},
	title = {{Highly efficient XUV generation via high-order frequency mixing}},
	journal = {New J. Phys.},
	volume = {22},
	number = {9},
	pages = {093030},
	year = {2020},
	month = {Sep},
	issn = {1367-2630},
	publisher = {IOP Publishing},
	doi = {10.1088/1367-2630/abae89}
}

@article{Strelkov_2005,
  title = {High-harmonic generation in a dense medium},
  author = {Strelkov, V. V. and Platonenko, V. T. and Becker, A.},
  journal = {Phys. Rev. A},
  volume = {71},
  issue = {5},
  pages = {053808},
  numpages = {8},
  year = {2005},
  month = {May},
  publisher = {American Physical Society},
  doi = {10.1103/PhysRevA.71.053808},
  url = {https://link.aps.org/doi/10.1103/PhysRevA.71.053808}
}

@article{Becker_1994,
  title = {Modeling harmonic generation by a zero-range potential},
  author = {Becker, W. and Long, S. and McIver, J. K.},
  journal = {Phys. Rev. A},
  volume = {50},
  issue = {2},
  pages = {1540--1560},
  numpages = {0},
  year = {1994},
  month = {Aug},
  publisher = {American Physical Society},
  doi = {10.1103/PhysRevA.50.1540},
  url = {https://link.aps.org/doi/10.1103/PhysRevA.50.1540}
}

@article{EVA,
  title = {Analytical solutions for strong field-driven atomic and molecular one- and two-electron continua and applications to strong-field problems},
  author = {Smirnova, Olga and Spanner, Michael and Ivanov, Misha},
  journal = {Phys. Rev. A},
  volume = {77},
  issue = {3},
  pages = {033407},
  numpages = {27},
  year = {2008},
  month = {Mar},
  publisher = {American Physical Society},
  doi = {10.1103/PhysRevA.77.033407},
  url = {https://link.aps.org/doi/10.1103/PhysRevA.77.033407}
}

@article{Frolov_2007,
  title = {Description of harmonic generation in terms of the complex quasienergy. I. General formulation},
  author = {Frolov, M. V. and Flegel, A. V. and Manakov, N. L. and Starace, Anthony F.},
  journal = {Phys. Rev. A},
  volume = {75},
  issue = {6},
  pages = {063407},
  numpages = {14},
  year = {2007},
  month = {Jun},
  publisher = {American Physical Society},
  doi = {10.1103/PhysRevA.75.063407},
  url = {https://link.aps.org/doi/10.1103/PhysRevA.75.063407}
}

@article{Cheng_2018,
  title = {Nonperturbative model for optical response under intense periodic fields with application to graphene in a strong perpendicular magnetic field},
  author = {Cheng, J. L. and Guo, C.},
  journal = {Phys. Rev. B},
  volume = {97},
  issue = {20},
  pages = {205406},
  numpages = {7},
  year = {2018},
  month = {May},
  publisher = {American Physical Society},
  doi = {10.1103/PhysRevB.97.205406},
  url = {https://link.aps.org/doi/10.1103/PhysRevB.97.205406}
}

@article{Ryabikin2023,
	author = {M. Yu. Ryabikin and M. Yu. Emelin and V. V. Strelkov},
	title = {Attosecond electromagnetic pulses: generation, measurement, and application. Attosecond metrology and spectroscopy},
	publisher = {Physics-Uspekhi},
	year = {2023},
	journal = {Phys. Usp.},
	volume = {66},
	number = {4},
	pages = {360-380},
	url = {https://ufn.ru/en/articles/2023/4/b/},
	doi = {10.3367/UFNe.2021.10.039078}
}

@article{Kostin2016,
  title = {Ionization-Induced Multiwave Mixing: Terahertz Generation with Two-Color Laser Pulses of Various Frequency Ratios},
  author = {Kostin, V. A. and Laryushin, I. D. and Silaev, A. A. and Vvedenskii, N. V.},
  journal = {Phys. Rev. Lett.},
  volume = {117},
  issue = {3},
  pages = {035003},
  numpages = {5},
  year = {2016},
  month = {Jul},
  publisher = {American Physical Society},
  doi = {10.1103/PhysRevLett.117.035003},
  url = {https://link.aps.org/doi/10.1103/PhysRevLett.117.035003}
}

@article{Birulia2024,
	author = {Birulia, V. A. and Khokhlova, M. A. and Strelkov, V. V.},
	title = {Macroscopic effects in generation of attosecond XUV pulses via high-order frequency mixing in gases and plasma},
	journal = {New J. Phys.},
	volume = {26},
	pages = {023005},
	year = {2024},
	doi = {10.1088/1367-2630/ad2130}
}

@article{Xiong_2017,
doi = {10.1088/1361-6455/50/3/032001},
url = {https://doi.org/10.1088/1361-6455/50/3/032001},
year = {2017},
month = {jan},
publisher = {IOP Publishing},
volume = {50},
number = {3},
pages = {032001},
author = {Xiong, Wei-Hao and Peng, Liang-You and Gong, Qihuang},
title = {Recent progress of below-threshold harmonic generation},
journal = {Journal of Physics B: Atomic, Molecular and Optical Physics}
}

@article{Antonov2025,
      title={Optimal conditions for the generation of moderate-order harmonics of a short-wave field by helium atoms}, 
      author={V. A. Antonov and I. R. Khairulin and M. Yu. Emelin and M. M. Popova and E. V. Gryzlova and M. Yu. Ryabikin},
       url={}, 
      journal = {Phys. Rev. A },
      year = {2025},
      volume = {111},
      issue = {},
      pages = {053502}, 
}

@article{data_nonperturbative,
doi = {},
url = {https://github.com/bondarenkosofiya/Nonperturbative-regime-of-low-order-harmonic-generation-in-intense-low-frequency-laser-field},
year = {2026},
month = {},
publisher = {},
volume = {},
number = {},
pages = {},
author = {Strelkov, V. V. and Bondarenko, S. A.},
title = {https://github.com/bondarenkosofiya/Nonperturbative-regime-of-low-order-harmonic-generation-in-intense-low-frequency-laser-field},
journal = {Github},
}

@article{Brunel,
author = {F. Brunel},
journal = {J. Opt. Soc. Am. B},
number = {4},
pages = {521--526},
publisher = {Optica Publishing Group},
title = {Harmonic generation due to plasma effects in a gas undergoing multiphoton ionization in the high-intensity limit},
volume = {7},
month = {Apr},
year = {1990},
url = {https://opg.optica.org/josab/abstract.cfm?URI=josab-7-4-521},
doi = {10.1364/JOSAB.7.000521},
}

@article{ABecker2015,
doi = {},
url = {},
year = {2015},
month = {},
publisher = {},
volume = {91},
number = {},
pages = {023402},
author = { Spott, A. and Becker, A. and Jaron-Becker, A.},
title = {Transition from perturbative to nonperturbative interaction in low-order-harmonic generation},
journal = {Phys Rev A},
}

@article{ABecker2017,
doi = {},
url = {},
year = {2017},
month = {},
publisher = {},
volume = {96},
number = {},
pages = {053404},
author = { Spott, A.  and Jaron-Becker, A. and Becker, A.},
title = {Time-dependent susceptibility of a helium atom in intense laser pulses},
journal = {Phys Rev A},
}

@article{ABecker2014,
doi = {10.1103/PhysRevA.90.013426},
url = {https://journals.aps.org/pra/abstract/10.1103/PhysRevA.90.013426},
year = {2014},
month = {},
publisher = {},
volume = {90},
number = {},
pages = {013426},
author = {Spott, A.  and Jaron-Becker, A. and Becker, A.},
title = {},
journal = {Phys Rev A},
}

@article{Emelin2025,
doi = {},
url = {},
year = {2025},
month = {},
publisher = {},
volume = {57},
number = {},
pages = {434},
author = {Emelin, M. Y. and Ryabikin, M. Y.},
title = {High-ellipticity resonant below-threshold harmonic generation by a helium atom driven by a moderately intense elliptically polarized laser field},
journal = {Optical and Quantum Electronics },
}

@article{Vvedenskii_2014,
doi = {10.1103/PhysRevLett.112.055004},
url = {https://journals.aps.org/prl/abstract/10.1103/PhysRevLett.112.055004},
year = {2014},
month = {},
publisher = {},
volume = {112},
number = {},
pages = {055004},
author = {Vvedenskii, N. V. and Korytin, A. I. and  Kostin,  V. A. and Murzanev, A. A. and Silaev, A. A. and Stepanov, A. N.},
title = {Two-color laser-plasma generation of terahertz radiation using a frequency-tunable half harmonic of a femtosecond pulse},
journal = {Phys Rev Lett },
}

@book{duffing,
  title={The Duffing equation: nonlinear oscillators and their behaviour},
  author={Kovacic, Ivana and Brennan, Michael J},
  year={2011},
  publisher={John Wiley \& Sons}
}

@article{volkova_nonlinear_2013,
	title = {Nonlinear polarization response of a gaseous medium in the regime of atom stabilization in a strong radiaion field},
	volume = {116},
	issn = {1090-6509},
	url = {https://doi.org/10.1134/S106377611303014X},
	doi = {10.1134/S106377611303014X},
	language = {english},
	number = {3},
	urldate = {2026-03-11},
	journal = {Journal of Experimental and Theoretical Physics},
	author = {Volkova, E. A. and Popov, A. M. and Tikhonova, O. V.},
	month = mar,
	year = {2013},
	pages = {372--380},
}

@article{Loriot_2009,
author = {V. Loriot and E. Hertz and O. Faucher and B. Lavorel},
journal = {Opt. Express},
number = {16},
pages = {13429--13434},
publisher = {Optica Publishing Group},
title = {Measurement of high order Kerr refractive index of major air components},
volume = {17},
month = {Aug},
year = {2009},
url = {https://opg.optica.org/oe/abstract.cfm?URI=oe-17-16-13429},
doi = {10.1364/OE.17.013429},
}

@article{Tolstikhin_2013,
  title = {Siegert Pseudo-States as a Universal Tool: Resonances, $\mathit{S}$ Matrix, Green Function},
  author = {Tolstikhin, Oleg I. and Ostrovsky, Valentin N. and Nakamura, Hiroki},
  journal = {Phys. Rev. Lett.},
  volume = {79},
  issue = {11},
  pages = {2026--2029},
  numpages = {0},
  year = {1997},
  month = {Sep},
  publisher = {American Physical Society},
  doi = {10.1103/PhysRevLett.79.2026},
  url = {https://link.aps.org/doi/10.1103/PhysRevLett.79.2026}
}

@article{Popov_2013,
doi = {10.1088/1612-2011/10/8/085303},
url = {https://doi.org/10.1088/1612-2011/10/8/085303},
year = {2013},
month = {jun},
publisher = {IOP Publishing},
volume = {10},
number = {8},
pages = {085303},
author = {Popov, A M and Tikhonova, O V and Volkova, E A},
title = {Polarization response of an atomic system in a strong mid-IR field},
journal = {Laser Physics Letters}
}

@article{Bejot_2010,
  title = {Higher-Order Kerr Terms Allow Ionization-Free Filamentation in Gases},
  author = {B\'ejot, P. and Kasparian, J. and Henin, S. and Loriot, V. and Vieillard, T. and Hertz, E. and Faucher, O. and Lavorel, B. and Wolf, J.-P.},
  journal = {Phys. Rev. Lett.},
  volume = {104},
  issue = {10},
  pages = {103903},
  numpages = {4},
  year = {2010},
  month = {Mar},
  publisher = {American Physical Society},
  doi = {10.1103/PhysRevLett.104.103903},
  url = {https://link.aps.org/doi/10.1103/PhysRevLett.104.103903}
}

@article{Mechain_2004,
doi = {10.1007/s00340-004-1557-8},
url = {https://doi.org/10.1007/s00340-004-1557-8},
year = {2004},
month = {},
publisher = {},
volume = {79},
number = {},
pages = {379–382},
author = {M\'echain, G. and Couairon, A. and   Andr\'e, Y.-B. and D’Amico, C. and Franco, M. and Prade, B.  and Tzortzakis, S. and Mysyrowicz, A. and Sauerbrey, R. }, 
title = {Long-range self-channeling of infrared laser pulses in air: a new propagation regime without ionization},
journal = {Appl. Phys. B},
}

@article{Berge_2007,
doi = {10.1088/0034-4885/70/10/R03},
url = {https://doi.org/10.1088/0034-4885/70/10/R03},
year = {2007},
month = {sep},
publisher = {},
volume = {70},
number = {10},
pages = {1633},
author = {Berge, L and Skupin, S and Nuter, R and Kasparian, J and Wolf, J-P},
title = {Ultrashort filaments of light in weakly ionized, optically transparent media},
journal = {Reports on Progress in Physics}
}

@article{Couairon_2000,
    author = {Couairon, A. and Berge, L.},
    title = {Modeling the filamentation of ultra-short pulses in ionizing media},
    journal = {Physics of Plasmas},
    volume = {7},
    number = {1},
    pages = {193-209},
    year = {2000},
    month = {01},
    issn = {1070-664X},
    doi = {10.1063/1.873794},
    url = {https://doi.org/10.1063/1.873794},
}

@article{Couairon_2007,
title = {Femtosecond filamentation in transparent media},
journal = {Physics Reports},
volume = {441},
number = {2},
pages = {47-189},
year = {2007},
issn = {0370-1573},
doi = {https://doi.org/10.1016/j.physrep.2006.12.005},
url = {https://www.sciencedirect.com/science/article/pii/S037015730700021X},
author = {Couairon, A.  and Mysyrowicz, A.}
}

@article{Chin_2012,
doi = {10.1134/S1054660X11190054},
url = {https://link.springer.com/article/10.1134/S1054660X11190054},
year = {2012},
month = {},
publisher = {},
volume = {22},
number = {},
pages = {1},
author = { Chin, S. L. and  Wang, T. -J. and  Marceau, C.  and  Wu, J. and   Liu, J. S. and Kosareva,  O. and  Panov, N. and Chen, Y. P.  and  Daigle, J. -F. and   Yuan, S. and  Azarm, A. and  Liu, W. W.  and  Seideman, T. and Zeng, H. P. and  Richardson,  M. and  Li R. and  Xu Z. Z.},
title = {Advances in intense femtosecond laser filamentation in air},
journal = {Laser Physics},
}

@article{Kandidov_2013,
doi = {10.3367/UFNe.0183.201302b.0133},
url = {https://ufn.ru/ufn13/ufn13_2/ufn132b.pdf},
year = {2013},
month = {},
publisher = {},
volume = {56},
number = {},
pages = {123},
author = {Chekalin, S. V. and  Kandidov, V. P.},
title = {From self-focusing light beams to femtosecond laser pulse filamentation},
journal = {Phys. Usp. },
}

@article{Mlejnek_1998,
author = {M. Mlejnek and E. M. Wright and J. V. Moloney},
journal = {Opt. Lett.},
number = {5},
pages = {382--384},
publisher = {Optica Publishing Group},
title = {Dynamic spatial replenishment of femtosecond pulses propagating in air},
volume = {23},
month = {Mar},
year = {1998},
url = {https://opg.optica.org/ol/abstract.cfm?URI=ol-23-5-382},
doi = {10.1364/OL.23.000382},
}

@article{Ettoumi_2010,
  title = {Spectral dependence of purely-Kerr-driven filamentation in air and argon},
  author = {Ettoumi, W. and B\'ejot, P. and Petit, Y. and Loriot, V. and Hertz, E. and Faucher, O. and Lavorel, B. and Kasparian, J. and Wolf, J.-P.},
  journal = {Phys. Rev. A},
  volume = {82},
  issue = {3},
  pages = {033826},
  numpages = {5},
  year = {2010},
  month = {Sep},
  publisher = {American Physical Society},
  doi = {10.1103/PhysRevA.82.033826},
  url = {https://link.aps.org/doi/10.1103/PhysRevA.82.033826}
}

@article{Volkova_2012,
doi = {10.1070/QE2012v042n08ABEH014878},
url = {https://doi.org/10.1070/QE2012v042n08ABEH014878},
year = {2012},
month = {aug},
publisher = {},
volume = {42},
number = {8},
pages = {680},
author = {Volkova, E A and Popov, Alexander M and Tikhonova, O V},
title = {Polarisation response of a gas medium in the field of a high-intensity ultrashort laser pulse: high order Kerr nonlinearities or plasma electron component?},
journal = {Quantum Electronics}
}

@article{Kosareva_2011,
author = {Olga Kosareva and Jean-Francois Daigle and Nikolay Panov and Tiejun Wang and Sima Hosseini and Shuai Yuan and Gilles Roy and Vladimir Makarov and See Leang Chin},
journal = {Opt. Lett.},
number = {7},
pages = {1035--1037},
publisher = {Optica Publishing Group},
title = {Arrest of self-focusing collapse in femtosecond air filaments: higher order Kerr or plasma defocusing?},
volume = {36},
month = {Apr},
year = {2011},
url = {https://opg.optica.org/ol/abstract.cfm?URI=ol-36-7-1035},
doi = {10.1364/OL.36.001035},
}

@article{Shipilo_2017,
author = {D. E. Shipilo and N. A. Panov and E. S. Sunchugasheva and D. V. Mokrousova and A. V. Shutov and V. D. Zvorykin and N. N. Ustinovskii and L. V. Seleznev and A. B. Savel'ev and O. G. Kosareva and S. L. Chin and A. A. Ionin},
journal = {Opt. Express},
number = {21},
pages = {25386--25391},
publisher = {Optica Publishing Group},
title = {Fifteen meter long uninterrupted filaments from sub-terawatt ultraviolet pulse in air},
volume = {25},
month = {Oct},
year = {2017},
url = {https://opg.optica.org/oe/abstract.cfm?URI=oe-25-21-25386},
doi = {10.1364/OE.25.025386},
}

@article{ Vrublevskaya_2023,
doi = {10.1134/S0021364023600301},
url = {https://doi.org/10.1134/S0021364023600301},
year = {2023},
month = {},
publisher = {},
volume = {117},
number = {6},
pages = {408–413},
author = {Vrublevskaya, N. R. and Shipilo, D. E.  and Nikolaeva , I. A. and  Panov, N. A. and Kosareva, O. G.},
title = {Nonlinear Response of Diluted Gases to an Ultraviolet Femtosecond Pulse},
journal = {JETP Letters},
}

@article{Nikolaeva_2026,
  title = {Second-harmonic generation in the air-based femtosecond plasma under loose focusing},
  author = {Nikolaeva, I. A. and Shipilo, D. E. and Rizaev, G. E. and Koribut, A. V. and Dick, T. A. and Pushkarev, D. V. and Levus, M. V. and Grudtsyn, Ya. V. and Vrublevskaya, N. R. and Panov, N. A. and Kosareva, O. G. and Seleznev, L. V. and Ionin, A. A.},
  journal = {Phys. Rev. E},
  volume = {113},
  issue = {2},
  pages = {025206},
  numpages = {11},
  year = {2026},
  month = {Feb},
  publisher = {American Physical Society},
  doi = {10.1103/x8lz-vwp5},
  url = {https://link.aps.org/doi/10.1103/x8lz-vwp5}
}

@article{Smetanin_2016,
title = {Role of coherent resonant nonlinear processes in the ultrashort KrF laser pulse propagation and filamentation in air},
journal = {Nuclear Instruments and Methods in Physics Research Section B: Beam Interactions with Materials and Atoms},
volume = {369},
pages = {87-91},
year = {2016},
issn = {0168-583X},
doi = {https://doi.org/10.1016/j.nimb.2015.10.032},
url = {https://www.sciencedirect.com/science/article/pii/S0168583X15010393},
author = {I.V. Smetanin and A.O. Levchenko and A.V. Shutov and N.N. Ustinovskii and V.D. Zvorykin}
}

@book{Elutin,
  title = {Quantum mechanics},
  publisher = {Nauka},
  year = {1976},
  author = {Elutin, P. V. and  Krivchenkov, V. D. },
  address = {Moscov},
  edition = {},
}

@article{Delone_1998,
	author = {N. B. Delone and V. P. Krainov},
	title = {Tunneling and barrier-suppression ionization of atoms and ions in a laser radiation field},
	publisher = {Physics-Uspekhi},
	year = {1998},
	journal = {Phys. Usp.},
	volume = {41},
	number = {5},
	pages = {469-485},
	url = {https://ufn.ru/en/articles/1998/5/c/},
	doi = {10.1070/PU1998v041n05ABEH000393}
    }

@article{Brodeur_1997,
author = {A. Brodeur and C. Y. Chien and F. A. Ilkov and S. L. Chin and O. G. Kosareva and V. P. Kandidov},
journal = {Opt. Lett.},
number = {5},
pages = {304--306},
publisher = {Optica Publishing Group},
title = {Moving focus in the propagation of ultrashort laser pulses in air},
volume = {22},
month = {Mar},
year = {1997},
url = {https://opg.optica.org/ol/abstract.cfm?URI=ol-22-5-304},
doi = {10.1364/OL.22.000304},
}

@article{Strelkov_UV_filamentation,
doi = {},
url = {},
year = {2026},
month = {},
publisher = {},
volume = {},
number = {},
pages = {},
author = {Strelkov, V.V and et al.},
title = {Ionizationless femtosecond {UV} pulses filamentation due to non-perturbative Kerr effect in transiently photoexcited molecules},
journal = {arXiv},
}

@article{Shipilo_2025,
  title = {Long-wavelength spectral shift in an ultraviolet filament},
  author = {Shipilo, D. E. and Vrublevskaya, N. R. and Nikolaeva, I. A. and Seleznev, L. V. and Pushkarev, D. V. and Rizaev, G. E. and Levus, M. V. and Ionin, A. A. and Panov, N. A. and Kosareva, O. G.},
  journal = {Phys. Rev. A},
  volume = {112},
  issue = {2},
  pages = {023516},
  numpages = {8},
  year = {2025},
  month = {Aug},
  publisher = {American Physical Society},
  doi = {10.1103/mgvk-wscf},
  url = {https://link.aps.org/doi/10.1103/mgvk-wscf}
}

@article{Telnov_2013,
  title = {Exterior complex scaling method in time-dependent density-functional theory: Multiphoton ionization and high-order-harmonic generation of Ar atoms},
  author = {Telnov, Dmitry A. and Sosnova, Ksenia E. and Rozenbaum, Efim and Chu, Shih-I},
  journal = {Phys. Rev. A},
  volume = {87},
  issue = {5},
  pages = {053406},
  numpages = {8},
  year = {2013},
  month = {May},
  publisher = {American Physical Society},
  doi = {10.1103/PhysRevA.87.053406},
  url = {https://link.aps.org/doi/10.1103/PhysRevA.87.053406}
}

@article{Romanov_2024,
  title = {Simulation of resonance-enhanced high-order harmonic generation and autoionization decay in ${\mathrm{Ga}}^{+}$ based on time-dependent density-functional theory},
  author = {Romanov, A. A. and Strelkov, V. V. and Silaev, A. A.},
  journal = {Phys. Rev. A},
  volume = {110},
  issue = {6},
  pages = {063109},
  numpages = {13},
  year = {2024},
  month = {Dec},
  publisher = {American Physical Society},
  doi = {10.1103/PhysRevA.110.063109},
  url = {https://link.aps.org/doi/10.1103/PhysRevA.110.063109}
}

@article{Hirata_2002,
    author = {Hirata, So and Ivanov, Stanislav and Grabowski, Ireneusz and Bartlett, Rodney J.},
    title = {Time-dependent density functional theory employing optimized effective potentials},
    journal = {The Journal of Chemical Physics},
    volume = {116},
    number = {15},
    pages = {6468-6481},
    year = {2002},
    month = {04},
   issn = {0021-9606},
    doi = {10.1063/1.1460869},
    url = {https://doi.org/10.1063/1.1460869},
}

@article{Yang_2002,
  title = {Direct Method for Optimized Effective Potentials in Density-Functional Theory},
  author = {Yang, Weitao and Wu, Qin},
  journal = {Phys. Rev. Lett.},
  volume = {89},
  issue = {14},
  pages = {143002},
  numpages = {4},
  year = {2002},
  month = {Sep},
  publisher = {American Physical Society},
  doi = {10.1103/PhysRevLett.89.143002},
  url = {https://link.aps.org/doi/10.1103/PhysRevLett.89.143002}
}

@book{press2007numerical,
  title={Numerical recipes 3rd edition: The art of scientific computing},
  author={Press, William H},
  year={2007},
  publisher={Cambridge university press}
}

@book{DFT,
  title={Time-Dependent Density-Functional Theory:
Concepts and Applications},
  author={Ullrich, C. A.},
  year={2012},
  publisher={Oxford University}
}

@article{Zhang_2024,
doi = {10.1038/s41586-024-07839-6},
url = {https://doi.org/10.1038/s41586-024-07839-6},
year = {2024},
month = {},
publisher = {},
volume = {633},
number = {},
pages = {63–70},
author = {Zhang, C. and Ooi, T. and  Higgins, J.S. and 
  Doyle, Jack F. and  and  von der Wense, Lars and Beeks, Kjeld and   Leitner, Adrian  and  Kazakov, Georgy A. and  and  Li, Peng and   Thirolf, Peter G. and  Schumm, Thorsten and Ye, Jun},
title = {Frequency ratio of the $^{229m}${Th} nuclear isomeric transition and the $^{87}${Sr} atomic clock},
journal = {Nature},
}

@article{Han_2025,
doi = {10.1038/s41586-025-09455-4},
url = {https://doi.org/10.1038/s41586-025-09455-4},
year = {2025},
month = {},
publisher = {},
volume = {645},
number = {},
pages = {95–100},
author = {Han, M. and Ji, JB. and Blech, A. and  Goetz, R. Esteban and  Allison, Cobin and  Greenman, Loren  and Koch,  Christiane P. and W\"orner, Hans Jakob},
title = {Attosecond control and measurement of chiral photoionization dynamics},
journal = {Nature},
}

@article{Wang_2026,
    author = {Wang, Bincheng and Lu, Bin and Fu, Yong and Tang, Xiangyu and Li, Baochang and Jin, Cheng},
    title = {Enhancement of Vacuum-Ultraviolet Laser Generation through Near-Threshold Harmonics Driven by a Few-Cycle Two-Color Laser Field},
    journal = {ACS Photonics},
    volume = {13},
    number = {4},
    pages = {958-970},
    year = {2026},
    month = {02},
    issn = {2330-4022},
    doi = {10.1021/acsphotonics.5c02138},
    url = {https://doi.org/10.1021/acsphotonics.5c02138},
    }

@article{Han_2026,
doi = {10.1038/s42005-026-02579-7},
url = {https://doi.org/10.1038/s42005-026-02579-7},
year = {2026},
month = {},
publisher = {},
volume = {9},
number = {},
pages = {145},
author = {Han, J. and Wang, B. and Tang, X. and Yin, Zhiming and Fu, Yong and   You, Jiahao and Li, Baochang and Wu, Shengfei and Jin, Cheng },
title = {Generation of vacuum ultraviolet vortex beams via near-threshold harmonics in argon gas driven by infrared Laguerre-Gaussian lasers},
journal = {Commun Phys},
}

@misc{Vabek2026,
      title={Modular multiscale approach to modelling high-harmonic generation in gases}, 
      author={Jan Vábek and Tadeáš Němec and Stefan Skupin and Fabrice Catoire},
      year={2026},
      eprint={2507.04115},
      archivePrefix={arXiv},
      primaryClass={physics.optics},
      url={https://arxiv.org/abs/2507.04115}, 
}

@article{Wang_2023,
  title = {Spatially resolved macroscopic near-threshold harmonics generated from harmoniclike and resonant transitions using an artificial neural network},
  author = {Wang, Bincheng and Wen, Tianyu and Fu, Yong and Li, Baochang and Wang, Kan and Jin, Cheng},
  journal = {Phys. Rev. A},
  volume = {108},
  issue = {5},
  pages = {053510},
  numpages = {8},
  year = {2023},
  month = {Nov},
  publisher = {American Physical Society},
  doi = {10.1103/PhysRevA.108.053510},
  url = {https://link.aps.org/doi/10.1103/PhysRevA.108.053510}
}

@article{Perelomov,
doi = {},
url = {},
year = {1966},
month = {},
publisher = {},
volume = {50},
number = {},
pages = {1393},
author = {Perelomov, A. M. and Popov, V. S. and Terent'ev, M. V.},
title = {Ionization of Atoms in an Alternating Electric Field [{Sov.} {Phys.} {JETP} 23, 924 (1966)]},
journal = {Zh. Eksp. Teor. Fiz.},
}

@article{Zhang_2025,
  title = {Spectral asymmetry effect in terahertz generation by two-color ionizing femtosecond pulses},
  author = {Zhang, Zhiwen and Kostin, Vasily A. and Silaev, Alexander A. and Liu, Zefu and Wang, Shixiang and Lu, Chenhui and Fan, Zhengquan and Liu, Yi},
  journal = {Phys. Rev. A},
  volume = {111},
  issue = {2},
  pages = {023508},
  numpages = {16},
  year = {2025},
  month = {Feb},
  publisher = {American Physical Society},
  doi = {10.1103/PhysRevA.111.023508},
  url = {https://link.aps.org/doi/10.1103/PhysRevA.111.023508}
}

\end{document}